\documentclass[
reprint,
superscriptaddress,
amsmath,amssymb,
prl,
nolongbibliography
]{revtex4-2}
 
\usepackage{mathtools}
\usepackage{graphicx}
\usepackage[colorlinks=true,linkcolor=blue,citecolor=blue,urlcolor=blue,pdfauthor={ },pdftitle={ },pdfsubject={ },pdfkeywords={ }]{hyperref}
\usepackage{times}
\usepackage[qm]{qcircuit}
\usepackage{braket}
\usepackage{enumitem}
\usepackage{dcolumn}
\usepackage{bm}

\usepackage{siunitx}
\usepackage{upgreek}

\usepackage{color}
\usepackage{tikz}
\usepackage{pgf}
\usetikzlibrary{spy}

\definecolor{dred}{rgb}{.8,0.2,.2}
\definecolor{ddred}{rgb}{.8,0.5,.5}
\definecolor{dblue}{rgb}{.2,0.2,.8}
\definecolor{dgreen}{rgb}{.2,0.5,.2}

\newcommand*{\tr}{\mathrm{tr}}

\newcommand*{\fin}{\mathrm{fin}}
\newcommand*{\ce}{\mathcal{E}}
\newcommand*{\cs}{\mathcal{S}}
\newcommand*{\cc}{\mathcal{C}}
\newcommand*{\cd}{\mathcal{D}}
\newcommand*{\qcircuit}{$\begin{matrix}\Qcircuit @C=2em @R=2em}
\newcommand*{\eqcircuit}{\end{matrix}$}

\graphicspath{{figures/}}

\newcommand*{\physus}{Department of Physics, Southern University of Science and Technology, Shenzhen 518055, China}
\newcommand*{\inssus}{Shenzhen Institute for Quantum Science and Engineering, Southern University of Science and Technology, Shenzhen 518055, China}

\begin{document}

\preprint{APS/123-QED}

\title{Experimental Validation of Enhanced Information Capacity by Quantum Switch in Accordance with Thermodynamic Laws}

\author{Cheng Xi}
\thanks{These authors contributed equally to this work.}
\affiliation{\physus}
\affiliation{Department of Physics, City University of Hong Kong, Tat Chee Avenue, Kowloon, Hong Kong SAR, China}

\author{Xiangjing Liu}
\thanks{These authors contributed equally to this work.}
\affiliation{\physus}

\author{Hongfeng Liu}
\affiliation{\physus}

\author{Keyi Huang}
\affiliation{\inssus}

\author{Xinyue Long}
\affiliation{\physus}
\affiliation{Quantum Science Center of Guangdong-Hong Kong-Macao Greater Bay Area, Shenzhen 518045, China}

\author{Daniel Ebler}
\affiliation{Theory Lab, Central Research Institute, 2012 Labs, Huawei Technologies Co. Ltd., Hong Kong SAR, China
}
\affiliation{Department of Computer Science, The University of Hong Kong, Pokfulam Road, Hong Kong SAR, China}

\author{Xinfang Nie}
\email{niexf@sustech.edu.cn}
\affiliation{\physus}
\affiliation{Quantum Science Center of Guangdong-Hong Kong-Macao Greater Bay Area, Shenzhen 518045, China}

\author{Oscar Dahlsten}
\email{oscar.dahlsten@cityu.edu.hk}
\affiliation{Department of Physics, City University of Hong Kong, Tat Chee Avenue, Kowloon, Hong Kong SAR, China}
\affiliation{\physus}
\affiliation{\inssus}
\affiliation{Institute of Nanoscience and Applications, Southern University of Science and Technology, Shenzhen 518055, China}

\author{Dawei Lu}
\email{ludw@sustech.edu.cn}
\affiliation{\physus}
\affiliation{\inssus}
\affiliation{Quantum Science Center of Guangdong-Hong Kong-Macao Greater Bay Area, Shenzhen 518045, China}
\affiliation{International Quantum Academy, Shenzhen 518055, China}

\date{\today}

\begin{abstract}
We experimentally probe the interplay of the quantum switch with the laws of thermodynamics. The quantum
switch places two channels in a superposition of orders and may be applied to thermalizing channels. Quantum-switching thermal channels has been shown to give apparent violations of the second law. Central to these apparent violations is how quantum switching channels can increase the capacity to communicate information. We experimentally show this increase and how it is consistent with the laws of thermodynamics, demonstrating how thermodynamic resources are consumed. We use a nuclear magnetic resonance approach with coherently controlled interactions of nuclear spin qubits. We verify an analytical upper bound on the increase in capacity for channels that preserve energy and thermal states, and demonstrate that the bound can be exceeded for an energy-altering channel. We show that the switch can be used to take a thermal state to a state that is not thermal, whilst consuming free energy associated with the coherence of a control system. The results show how the switch can be incorporated into
quantum thermodynamics experiments as an additional resource

\end{abstract}

\maketitle

\emph{\bfseries Introduction.}---The laws of thermodynamics are foundational principles in physics~\cite{lieb1999physics,landau2013statistical,liu2024inferring}. The first law is a variant of energy conservation. The second law can be stated in several, arguably equivalent, formulations~\cite{lieb1999physics,landau2013statistical}. One version is that the {\em free energy} of a system cannot increase~\cite{liu,lostaglio2015description}. The free energy captures the thermodynamic resource value of a system~\cite{RevModPhys.91.025001,lieb1999physics, Fernando2013resource}. Challenges against the second law have, whilst sharpening our understanding of the second law, to date, failed to stand up to close scrutiny~\cite{maruyama2009physics,bennett1982thermodynamics,szilard1929entropieverminderung,PhysRevLett.128.090602,doi:10.1126/science.1078955,Parrondo2015}. 
\begin{figure}[t!]
\centering
\includegraphics[width=1\linewidth]{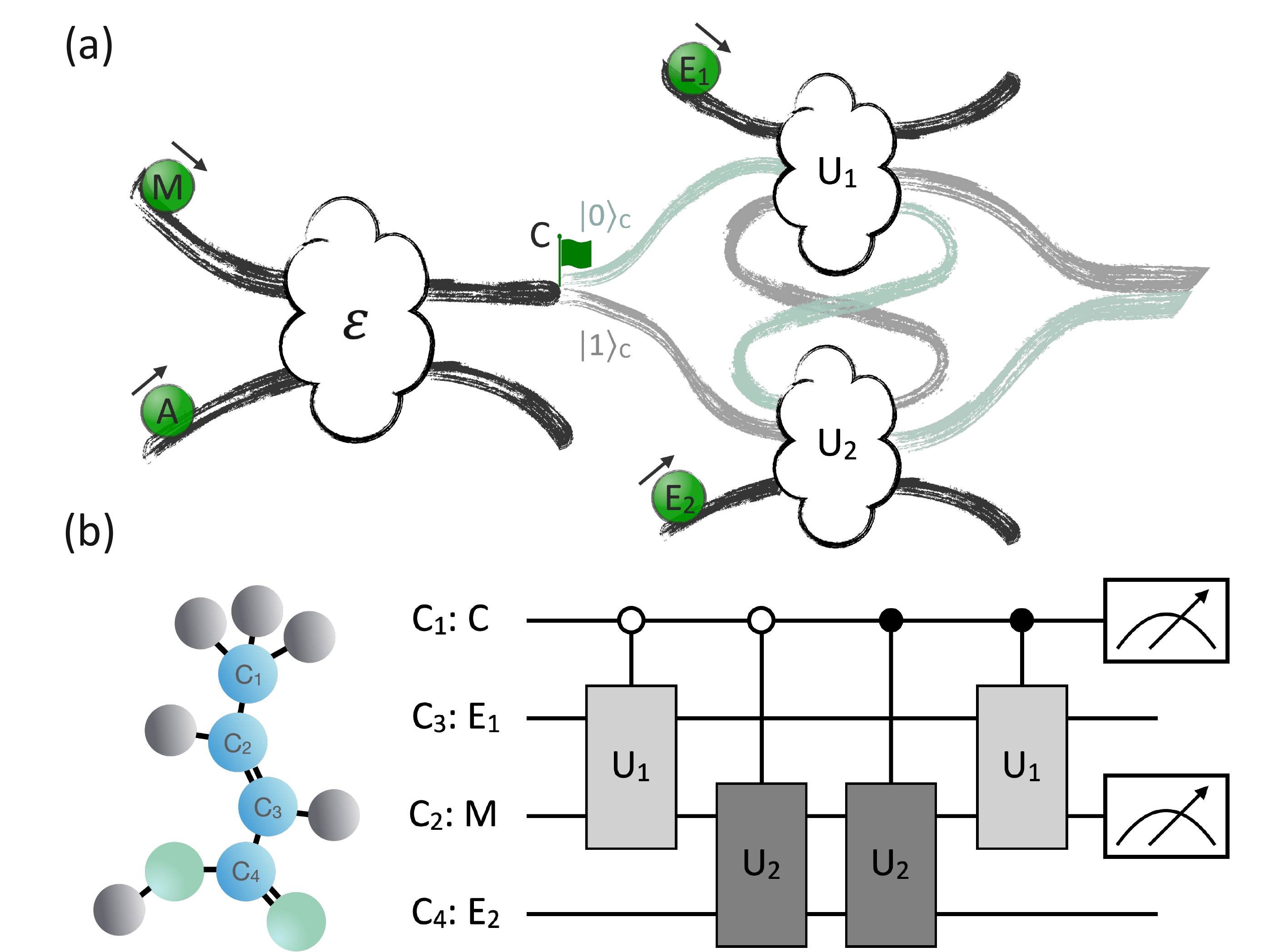}
\caption{{\bf The setup.} (a) A classical message stored in $A$ is encoded in qubit $M$ via the map $\varepsilon$. From the green starting flag, $M$ passes through unitary channels $U_1$ and $U_2$ coupling the system to thermalized qubits $E_1$ and $E_2$ respectively. $U_1$ and $U_2$ are applied in a superposition of orders. The order is controlled by qubit $C$: $\ket{0}_C$ means the green path is followed wherein $U_1$ is applied before $U_2$ and $\ket{1}_C$ means the gray path is followed and $U_2$ is applied before $U_1$. Thus $\ket{+}_C$ implies a superposition of the two orders. 
(b) Molecular structure of $^{13}$C transcrotonic acid, with the carbon nuclei spins involved in our experiment labeled. The quantum circuit implementing scenario (a) above is also shown.}
\label{setup}
\end{figure}

A recent challenge to the second law involves putting two consecutive channels in a superposition of orders with the application of a {\em quantum switch} to the channels (see Fig.~\ref{setup}(a))~\cite{PhysRevA.88.022318,PhysRevLett.121.090503,doi:10.1126/sciadv.1602589,PhysRevA.101.012340}. Such indefinite causal orders of channels have been shown to yield advantages for a variety of information related tasks~\cite{guerin2016exponential,renner2022computational,ebler2018enhance,zhao2020quantum,yin2023experimental,liu2023experimentally,liu2023unification}. The central role played by information and entropy in thermodynamics then suggests the possibility of similar thermodynamical advantages~\cite{PhysRevLett.125.070603,zhu2023charging}.

Applying the quantum switch to thermal channels leads to a larger retention of free energy. Thermalization removes information about the past, increasing the entropy and thereby lowering the free energy. How much information is retained can be quantified via the communication rate: the number of bits retained per physical qubit undergoing the evolution. The communication rate maximized over all encodings is known as the channel's information  capacity~\cite{PhysRevA.101.012340,cover1999elements,wilde2013quantum}. The information capacity of two switched channels can be larger than that of two channels applied in definite order~\cite{ebler2018enhance,salek2018quantum,8966996}. 
This retention of information about the initial state means a lower loss of free energy. 

The thermodynamics of the quantum switch have accordingly been investigated~\cite{Simonov2022work,verma2024measuring,dieguez2023thermal,Guha2020thermodynamics,capela2023reassessing,felce2020quantum,zhu2023charging}. Switched thermal channels allow for the separation of hot and cold, creating further tension with the second law~\cite{felce2020quantum,PhysRevLett.129.100603,cao2022quantum, PhysRevLett.129.100603}. Theoretical analysis showed that switching thermal channels in fact consumes a thermodynamic resource, and is consistent with the first and second laws of thermodynamics, as long as the increase in information capacity is strictly bounded~\cite{liu}. These results motivate investigating the gains and resource consumption of employing the switch in quantum thermodynamical experiments.

We employ the NMR platform for that purpose. We conduct 4-qubit experiments using an ensemble of nuclear spins via the NMR technique, as depicted in Fig.~\ref{setup}(b). The NMR platform has proven to be a powerful tool for the experimental study of quantum thermodynamics~\cite{PhysRevLett.129.100603,Micadei2019,PhysRevLett.123.240601,VIEIRA2023100105}. NMR is particularly suitable for implementing switched channels given the possibility to implement the controlled 2-qubit interactions in Fig.~\ref{setup}(b).

We show experimentally that the information capacity increase indeed consumes a thermodynamic resource: the free energy of coherence associated with the control qubit. The increase respects a highly restrictive analytical bound for quantum switched thermalisations, in agreement with the second law of thermodynamics and energy conservation. We also demonstrate that the same bound can in fact be violated by an operation that leaves the thermal state invariant but changes the energy, showing that additional thermal resources can lead to a greater increase in information capacity under the switch. The experiments demonstrate how the switch can be employed experimentally as an additional resource in quantum thermodynamics.

\emph{\bfseries Initial state and dynamics.}---The experiment aims to implement the interactions depicted in Fig.~\ref{setup}. 

The initial joint state of the two-level systems  $A$, $M$, $C$, $E_1$ and $E_2$ depicted in Fig.~\ref{setup}(a) is as follows. The initial state of $AM$ is 
$\rho_{AM}=p \ket{00}_{AM}\bra{00}+(1-p)\ket{11}_{AM}\bra{11},$ and unless otherwise stated $p=1/2$. $A$ is a classical record of the message recorded in $M$. Heat bath qubits $E_1$ and $E_2$ are initially in thermal states $\tau = e^{-H/kT}/\tr\left(e^{-H/kT}\right)$ where $T$ is the temperature, $k$ Boltzmann's constant and, for experimental simplicity, $H_{E_j}=H_M=H_C=-\sigma_z$ for $j=1,2$. The state of the switch control system $C$ is initially $\sigma_C$ (more details below). The total initial state we aim to realise is $ \sigma_C \otimes\rho_{AM}\otimes \tau_{E_1}\otimes \tau_{E_2}$.

The time evolution without the switch is a sequence of unitary interactions $U_1$ and $U_2$ between $M$ and the heat bath qubits $E_1$ and $E_2$ respectively. $U_j$ is said to satisfy the energy conservation condition if $[U_j, H_{E_j}+H_M] =0$. Energy conservation, which is closely connected to the first law, necessitates a specific form for $U_j$~\cite{scarani2002thermalizing,kraus2001optimal,liu} which is essentially that $U_j(\theta)=  e^{i\theta S_j }= \cos(\theta)\,\mathbb{I}+i\sin(\theta) \, S_j$, where $S_j$ represents the swap operator between $M$ and $E_j$ (see Sec.~S6 of SM~\cite{supp}\nocite{Xin_2018,RevModPhys.76.1037,JONES201191,KHANEJA2005296,park2016simulation}). The parameter $s= \sin \theta  \in[0,1]$ can by inspection be viewed as the \textit{thermalization strength}. Under each unitary the local dynamics of $M$ is then described by a map $\ce_j(\rho_M) =\tr_{E}\left(U_j (\rho_M\otimes\tau_{E_j}) U_j^\dag\right)$, where $\rho_M$ is the state of system $M$.

The \textit{quantum switch} $\cs$ by definition superposes the orders of any two consecutive channels, namely $\cc_2\circ\cc_1$ and $\cc_1\circ\cc_2$~\cite{PhysRevA.88.022318,PhysRevLett.101.060401,ebler2018enhance,PhysRevLett.124.190503}. Here $\cc_1$ and $\cc_2$ are unitary and the dynamics under the quantum switch is described by the overall unitary 
\begin{equation}
L=\ket{0}_C\bra{0}\otimes U_2U_1+\ket{1}_C\bra{1}\otimes U_1U_2.\label{l}
\end{equation}
If $U_1$ and $U_2$ are energy conserving $L$ is energy conserving~\cite{liu}.  When the reduced state on $C$, $\sigma_C =\op{0}{0}$, the order of the two interactions remains well-defined. When $\sigma_C =\op{+}{+}$, the order is maximally superposed. As mentioned above we initialise $C$ in the state of $\sigma_C=\lambda\op{+}{+}+(1-\lambda)\op{0}{0}$ where $\lambda\in[0,1]$ describes to what extent the quantum switch is on (a superposition of $\ket{0}$ and $\ket{+}$ yields similar conclusions~\cite{liu}). The total final state we aim to realise is $\left[\mathbb{I}_A\otimes L\right](\sigma_C \otimes \rho_{AM}\otimes \tau_{E_1}\otimes \tau_{E_2})\left[\mathbb{I}_A\otimes L^{\dagger}\right]$.

\begin{figure}[t]
\centering
\includegraphics[width=1\linewidth]{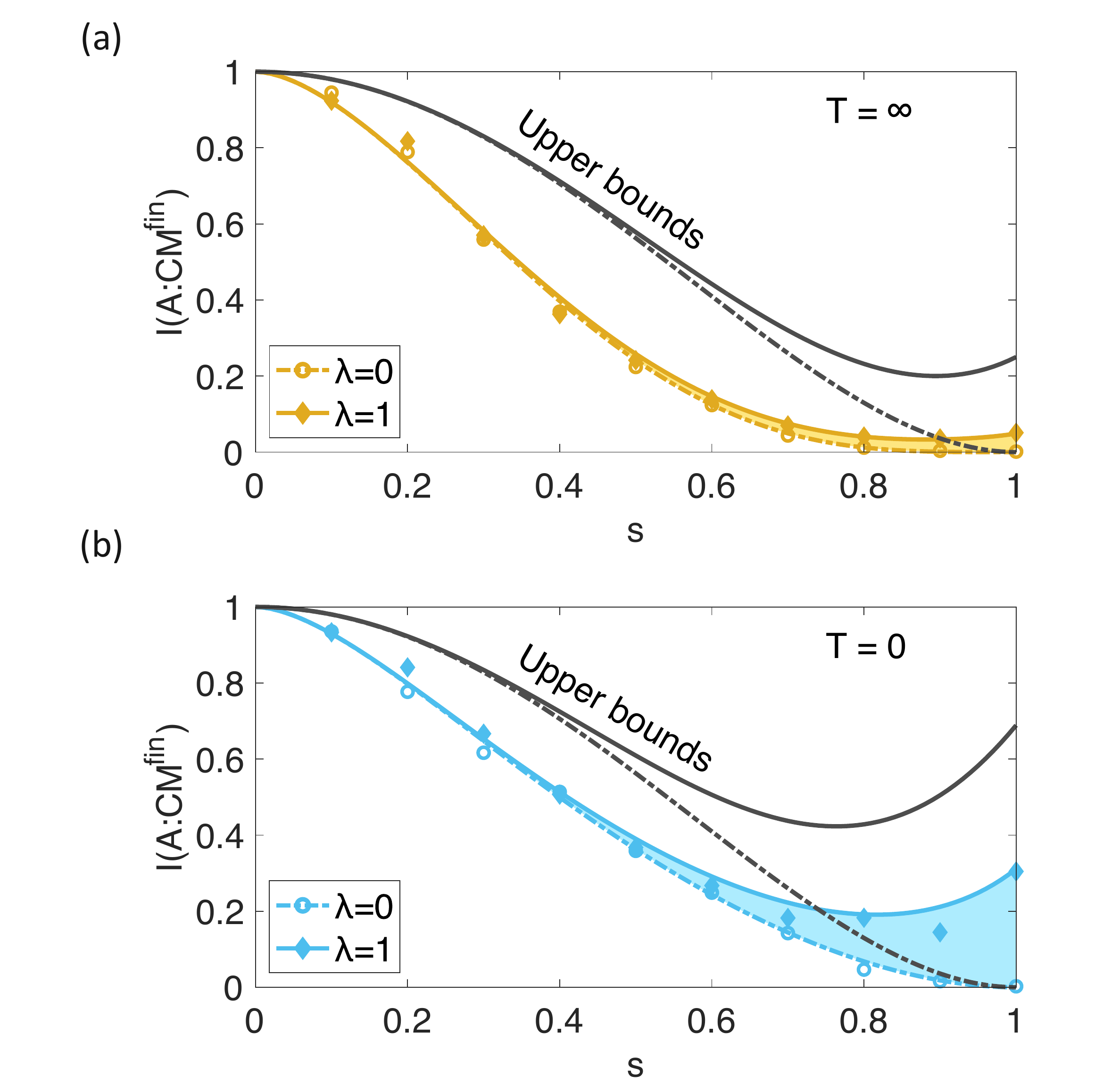}
\caption{\textbf{Bounded mutual information increase from switch.} The final mutual information $I(A:CM^\mathrm{fin})$ decreases with increasing thermalisation strength $s$. We verify experimentally that if the switch is ON ($\lambda=1$, diamonds) the decay is lower than if it is OFF ($\lambda=0$, circles). Analytical upper bounds~\cite{liu} (see Sec.~S2 of SM~\cite{supp}) for $I(A:CM^\mathrm{fin})$ are verified. For large $s$ some data points (diamonds) are above the upper bound for the case of the switch being OFF, showing a provable advantage from employing the switch. The standard deviations of the measurement outcomes are omitted because they do not exceed the size of the circles and diamonds (see Sec.~S11 of SM~\cite{supp}).
}
\label{main}
\end{figure}

\begin{figure*}[t]
\centering
\includegraphics[width=1\linewidth]{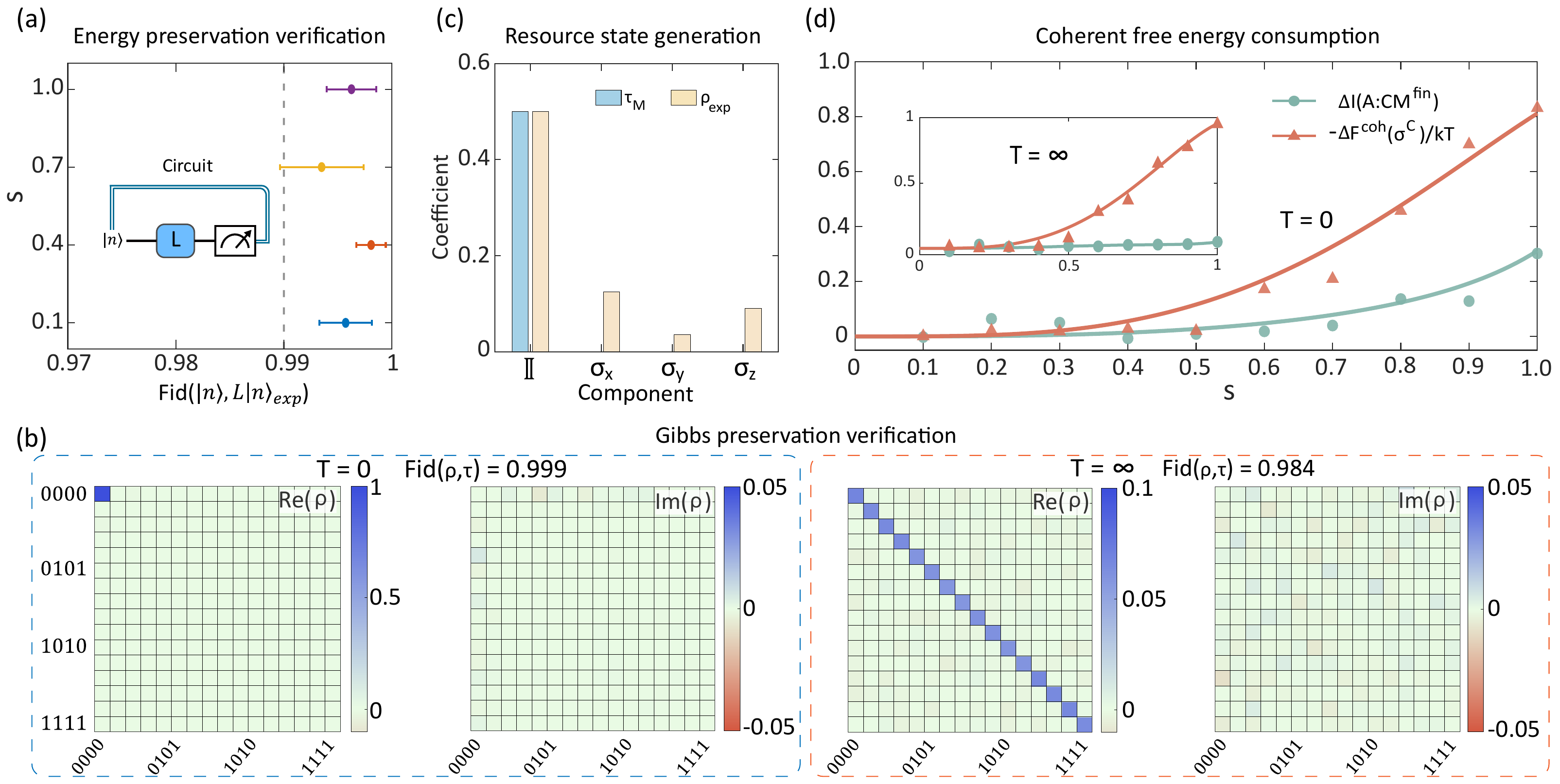}
\caption{{\bf Verification that quantum switch is a thermodynamic resource and consumes free energy of coherence.} (a) {\em The overall unitary $L$ conserves energy.} Fidelities between input and output energy eigenstates $\ket{n}_{CME_1E_2}$ ($n=0,\ldots,15$) are shown for varying thermalization strengths $s$. The average input-output fidelity of the eigenstates is close to 1 with a small standard deviation, verifying that $L$ essentially leaves all the energy eigenstates invariant. (b) {\em $L$ conserves Gibbs states.} The final states after $L$, given initial Gibbs states are shown. The fidelities to the theoretical Gibbs state are very close to 1. The thermalisation strength $s=1$. This demonstrates Gibbs state invariance under $L$.  (c) {\em Switch is a thermodynamic resource.} Combining the switch with the thermodynamically free swap unitary $S_{CM}$ leads to a map that is {\em not} Gibbs preserving, as described around Eq.~\eqref{eq:notfree}.
The density matrices of the thermal state $\tau_M$ (blue bars) and the experimental outcome (orange bars) for $s=1$ and $T=\infty$ are shown. The states are evidently different with a trace distance of 0.154.
(d) {\em Switch consumes free energy of coherence.} Increase in information capacity and the free energy of coherence of $C$ cost (normalized by $kT$) are presented as functions of $s$. The observed relation is consistent with the theoretical analysis showing that the free energy of coherence is consumed~\cite{liu}.}
\label{free}\label{free_energy}
\end{figure*}

\emph{\bfseries Thermodynamical and information theoretic quantities.}---We use the resource theory paradigm of states and operations being either {\em free} or {\em resources}. We call Gibbs states at the ambient temperature $T$ \textit{free} and other states, which are not in equilibrium and have higher free energy, as \textit{resources}~\cite{lieb1999physics}.
So-called \textit{thermal operations}~\cite{janzing2000thermodynamic} that are energy preserving and Gibbs state preserving ($\ce(\tau)=\tau$) for the given ambient temperature $T$, are considered free. The free energy $F(\rho) = \tr(\rho H) - kTS(\rho)$ where $S(\cdot)$ is the von Neumann entropy, quantifies the resource value of the system state in question, and decreases monotonically under free operations under the second law (see Sec.~S7 of SM~\cite{supp}):
\begin{equation} \label{eq:2ndlaw}
\Delta F(\rho) = F(\ce (\rho))-F(\rho) \leq 0. 
\end{equation}

We will investigate the switch-induced increase of the final mutual information between the record $A$ and $CM$, the output of the channel and switch. $C$ is included because the experimenter will have access to $C$ at the end. The mutual information is defined as $I(A:CM)=S(A)-S(A|CM)=S(\rho_{A})+S(\rho_{CM})-S(\rho_{ACM})$. We denote the final mutual information, after the dynamics of Eq.~\ref{l}, as $I(A:CM^\text{fin})$. The quantity $I(A:CM^\text{fin})$ (maximized over encodings) can be interpreted as a bound on the classical communication capacity of the quantum channel~\cite{wilde2013quantum}.


\emph{\bfseries Experimental setup.}---%
For experimental realization, we employ a nuclear spin system, where four distinct two-level $^{13}$C nuclei, arranged in a chain-like molecular structure~\cite{PhysRevLett.129.070502,PhysRevLett.126.110502,cheng2023noisy}, represent $C$, $M$, $E_1$, and $E_2$, respectively, as shown in Fig.~\ref{setup}(b). System $A$ is classical data that remains classically correlated with the quantum systems and is therefore for simplicity not stored in a qubit but rather as classical data. We perform the experiments with $M$ in both states $\ket{0}$ and $\ket{1}$, then combine the results from the two experiments to get results when system $A$ is included. The experiments are conducted at room temperature using a Bruker \qty[mode=text]{300}{MHz} NMR spectrometer. The Hamiltonian of the system is defined as $\mathcal{H}_\text{NMR}=-\pi\sum_i\nu_i\sigma_z^i+\pi\sum_{i<j}J_{ij}\sigma_z^i\sigma_z^j/2$, where $\sigma_z^i$ denotes the Pauli matrix $\sigma_z$ of the $i$-th spin, $\nu_i$ is the Larmor frequency, and $J_{ij}$ is the inter-spin coupling strength. The specific values of $\nu_i$ and $J_{ij}$ are listed in the Sec.~S1 of SM~\cite{supp}.

We can execute single-qubit rotations through transverse radio-frequency pulses and two-qubit rotations together with free evolution under $\mathcal{H}_\text{NMR}$, enabling the implementation of universal unitary operations, including $L$~\cite{supp}. This includes applying decoupling sequences to remove pair-wise interactions~\cite{supp}. As the pulse sequence time is much shorter than the relaxation times $T_2$ for $^{13}$C spins, the effect of decoherence can be neglected (see Sec.~S8 of SM~\cite{supp}).


\emph{\bfseries Experimental results.}---%
Our experimental results can be grouped into subsections: (I) Bounded mutual information increase from switch, (II) Verification that quantum switch is a thermodynamic resource, (III) Resource boost increases information capacity activation.

\emph{Results I: Bounded mutual information increase from switch:} We investigate the quantum mutual information \mbox{$I(A:CM^{\fin})$} between the record $A$ and the output $CM^{\text{fin}}$ under the switch. Our experimental circuit is illustrated in Fig.~\ref{setup}(b). The output state of $CM$ is measured to obtain its mutual information with the record $A$.  We probe several values of thermalization strength $s\in [0,1]$ and two extremes of temperature. The experimental results are presented in Fig.~\ref{main}, showing an increase in mutual information when the quantum switch is activated. Moreover, the increase remains constrained, never exceeding the theoretical upper bound derived from demanding that energy conservation and the second law apply to the overall unitary $L$~\cite{liu}. This experimentally demonstrates a mutual information increase from the switch in consistency with the laws of thermodynamics.

\emph{Results II: Verification that quantum switch is a thermodynamic resource:} We show, in three steps, that the switch is not a thermodynamically free operation.

Firstly, we verify that the overall unitary $L$ which is part of the switch action, is a free operation. We vary the thermalization strength $s$ with certain illustrative $s$ values used for the data being shown in Fig.~\ref{free} and the rest in Sec.~S3 of SM~\cite{supp}. We show $L$ is energy preserving in Fig.~\ref{free}(a). We show $L$ leaves the Gibbs state invariant in Fig.~\ref{free}(b). One should not conclude from this experiment that the switch should be viewed as thermodynamically free since the possible cost of the change in the control system has not been accounted for yet. 

Secondly, we show that the overall switch action, if combined with a particular free operation is actually {\em not} free. Ignoring the heat bath, the action of the switched channels on $M$ is given by $\mathcal{S}_{\sigma_C}(\ce_1, \ce_2 ) (\rho_M)=\tr_E \left(L(\sigma_C \otimes \rho_M \otimes \tau_{E_1} \otimes \tau_{E_2} ) L^\dag \right)$, where $\sigma_C$ is the state of the control system. In Fig.~\ref{free}(c), setting $s=1$, $T=\infty$, and $S_{\sigma_C}=S_{\ket{+}}$,  we have experimentally observed that 
\begin{equation}
\label{eq:notfree}
\tr_C\left[S_{CM}\circ\cs_{\ket{+}}(\ce_1,\ce_2)(\tau_M)\right]\neq\tau_M
\end{equation}
where $S_{CM}$ is the unitary swap between $C$ and $M$. $S_{CM}$ should be free as it is energy preserving and Gibbs preserving ($S_{CM}(\tau\otimes \tau) S_{CM}^{\dagger}=\tau\otimes \tau$) by inspection. Inequality \eqref{eq:notfree} shows that $\tr_C \circ S_{CM}\circ\cs_{\ket{+}}(\ce_1,\ce_2)$ is not free, as it does not preserve Gibbs states. Therefore, the switched channel $\cs_{\ket{+}}(\ce_1,\ce_2)$ must be not free.

Thirdly, we show that the quantum switch specifically consumes a thermodynamic resource termed the \textit{free energy of coherence} associated with $\sigma_C$. The free energy of coherence for any state $\rho$ can be defined as $F_\mathrm{coh}(\rho)=F(\rho)-F(\cd_H(\rho))$, where $\cd_H(\cdot)$ represents the operator that kills the off-diagonal elements of $\rho$ in the energy eigenbasis~\cite{lostaglio2015description,Baumgratz2014quantifying,Winter2016operational}. We experimentally quantified both the increase in mutual information and the cost of free energy of coherence of $C$, as illustrated in Fig.~\ref{free_energy}(d). The observed relation is consistent with the free energy of coherence being consumed, as derived in Ref.~\cite{liu}. 

\begin{figure}[t]
\centering
\includegraphics[width=.5\textwidth]{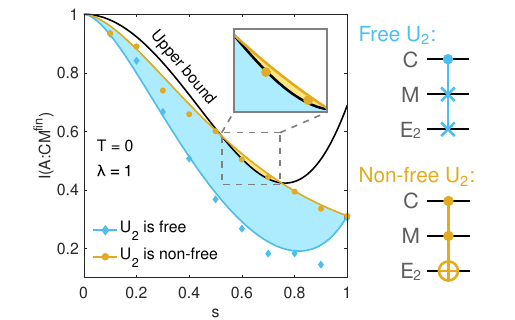}
\caption{{\bf Resource boost yields further information activation.} The final mutual information $I(A:CM^{\mathrm{fin}})$ 
is plotted versus the thermalisation strength $s$. $U_1$, a thermodynamically free partial swap matrix, and $U_2$ undergo a quantum switch. $U_2$ may equal $U_1$ or be a non-free partial CNOT (as depicted on the right). The partial CNOT choice is demonstrated to consistently yield higher mutual information, even violating the analytical upper bound~\cite{liu} on $I(A:CM^{\mathrm{fin}})$ for free $U_2$ around $s=\sin(\theta)=0.6$, as shown in the inset. This violation is consistent with how the partial CNOT is not free.}
\label{cnot}
\end{figure}

\emph{Results III: Resource boost increases information capacity activation:} We also examine a scenario in which additional resources are involved. We set $U_2$ to be a \textit{partial CNOT}, i.e.,
\begin{equation}
\label{eq:U2CNOT}
U_2(\theta)=e^{i\theta\cdot \text{CNOT}}.
\end{equation}
This choice of $U_2$ is interesting because it preserves Gibbs states ($U_2(\tau_M\otimes\tau_{E_2})U_2^\dag=\tau_M\otimes\tau_{E_2}$), but does not commute with the Hamiltonian of the systems ($[U_2, H_{E_2}+H_M]\neq 0$). That means that the second law of thermodynamics is obeyed whereas energy is altered. This alteration allows for the existence of additional resources in the system~\cite{Lostaglio2015quantum,cwiklinski2015limitations}. 

We demonstrate that the mutual information can be further enhanced from the switch in this case. As presented in Fig.~\ref{cnot}, for the $U_2$ of Eq.~\ref{eq:U2CNOT}, the mutual information \mbox{$I(A:CM^\fin)$} can,  within specific parameter regimes, surpass the theoretical upper bound of $I(A:CM^\fin)$ for the case where both $U_1$ and $U_2$ are partial SWAPs~\cite{liu}. Fig.~\ref{cnot} shows the case of $T=0$. A qualitatively similar outcome is observed in the scenario of $T=\infty$ as shown in Sec.~S4 of SM~\cite{supp} (see Secs S1, S2 in SM~\cite{supp} for details on the temperature $T$).

This increase in final mutual information can be directly attributed to the increase in energy available (see Sec.~S5 of SM~\cite{supp}). The results moreover show that the thermodynamic advantage in free energy conservation of $M$ afforded by the switch can be enhanced further than in previously known examples~\cite{liu,ebler2018enhance}.

\emph{\bfseries Conclusion and Outlook.}---%
The results show that the switch can be viewed as expanding the set of operations that one normally considers in quantum thermodynamics. Importantly the switch should be considered a resource rather than a free operation. The results point the way to a generalized resource theory of thermodynamics, where the quantum switch and potentially other indefinite causal structures are considered as resources. It should moreover be investigated further what kind of protocols the switch enables. For example,  mutual information can be converted into work via feedback control ~\cite{szilard1929entropieverminderung,toyabe2010information,sagawa2008second, PhysRevX.5.041011, rio2011thermodynamic} so novel work extraction protocols may be enabled by the increase in mutual information associated with the switch.

\emph{\bfseries Acknowledgements.}---This work is supported by the National Key Research and Development Program of China (2019YFA0308100), and National Natural Science Foundation of China (Grants No.12104213, 12075110, 12204230, 12050410246, 1200509, 12050410245), Science, Technology and Innovation Commission of Shenzhen Municipality (JCYJ20200109140803865),  Guangdong Innovative and Entrepreneurial Research Team Program (2019ZT08C044), Guangdong Provincial Key Laboratory (2019B121203002), City University of Hong Kong (Project No.~9610623), The Pearl River Talent Recruitment Program (2019QN01X298), and Guangdong Provincial Quantum Science Strategic Initiative (GDZX2303001 and GDZX2200001).

\emph{\bfseries Note.}---During the completion of our manuscript, we became aware of an independent experiment in optical setup by Tang et al. \cite{Tang2024demonstration}, which was concurrently posted to arXiv with ours.

\bibliography{bib}

\begin{thebibliography}{64}%
\makeatletter
\providecommand \@ifxundefined [1]{%
 \@ifx{#1\undefined}
}%
\providecommand \@ifnum [1]{%
 \ifnum #1\expandafter \@firstoftwo
 \else \expandafter \@secondoftwo
 \fi
}%
\providecommand \@ifx [1]{%
 \ifx #1\expandafter \@firstoftwo
 \else \expandafter \@secondoftwo
 \fi
}%
\providecommand \natexlab [1]{#1}%
\providecommand \enquote  [1]{``#1''}%
\providecommand \bibnamefont  [1]{#1}%
\providecommand \bibfnamefont [1]{#1}%
\providecommand \citenamefont [1]{#1}%
\providecommand \href@noop [0]{\@secondoftwo}%
\providecommand \href [0]{\begingroup \@sanitize@url \@href}%
\providecommand \@href[1]{\@@startlink{#1}\@@href}%
\providecommand \@@href[1]{\endgroup#1\@@endlink}%
\providecommand \@sanitize@url [0]{\catcode `\\12\catcode `\$12\catcode
  `\&12\catcode `\#12\catcode `\^12\catcode `\_12\catcode `\%12\relax}%
\providecommand \@@startlink[1]{}%
\providecommand \@@endlink[0]{}%
\providecommand \url  [0]{\begingroup\@sanitize@url \@url }%
\providecommand \@url [1]{\endgroup\@href {#1}{\urlprefix }}%
\providecommand \urlprefix  [0]{URL }%
\providecommand \Eprint [0]{\href }%
\providecommand \doibase [0]{https://doi.org/}%
\providecommand \selectlanguage [0]{\@gobble}%
\providecommand \bibinfo  [0]{\@secondoftwo}%
\providecommand \bibfield  [0]{\@secondoftwo}%
\providecommand \translation [1]{[#1]}%
\providecommand \BibitemOpen [0]{}%
\providecommand \bibitemStop [0]{}%
\providecommand \bibitemNoStop [0]{.\EOS\space}%
\providecommand \EOS [0]{\spacefactor3000\relax}%
\providecommand \BibitemShut  [1]{\csname bibitem#1\endcsname}%
\let\auto@bib@innerbib\@empty
\bibitem [{\citenamefont {Lieb}\ and\ \citenamefont
  {Yngvason}(1999)}]{lieb1999physics}%
  \BibitemOpen
  \bibfield  {author} {\bibinfo {author} {\bibfnamefont {E.~H.}\ \bibnamefont
  {Lieb}}\ and\ \bibinfo {author} {\bibfnamefont {J.}~\bibnamefont
  {Yngvason}},\ }\href {https://doi.org/10.1016/S0370-1573(98)00082-9}
  {\bibfield  {journal} {\bibinfo  {journal} {Phys. Rep.}\ }\textbf {\bibinfo
  {volume} {310}},\ \bibinfo {pages} {1} (\bibinfo {year} {1999})}\BibitemShut
  {NoStop}%
\bibitem [{\citenamefont {Landau}\ and\ \citenamefont
  {Lifshitz}(2013)}]{landau2013statistical}%
  \BibitemOpen
  \bibfield  {author} {\bibinfo {author} {\bibfnamefont {L.~D.}\ \bibnamefont
  {Landau}}\ and\ \bibinfo {author} {\bibfnamefont {E.~M.}\ \bibnamefont
  {Lifshitz}},\ }\href@noop {} {\emph {\bibinfo {title} {Statistical Physics:
  Volume 5}}},\ Vol.~\bibinfo {volume} {5}\ (\bibinfo  {publisher} {Elsevier},\
  \bibinfo {year} {2013})\BibitemShut {NoStop}%
\bibitem [{\citenamefont {Liu}\ \emph {et~al.}(2024{\natexlab{a}})\citenamefont
  {Liu}, \citenamefont {Chen},\ and\ \citenamefont
  {Dahlsten}}]{liu2024inferring}%
  \BibitemOpen
  \bibfield  {author} {\bibinfo {author} {\bibfnamefont {X.}~\bibnamefont
  {Liu}}, \bibinfo {author} {\bibfnamefont {Q.}~\bibnamefont {Chen}},\ and\
  \bibinfo {author} {\bibfnamefont {O.}~\bibnamefont {Dahlsten}},\ }\href
  {https://doi.org/10.1103/PhysRevA.109.032219} {\bibfield  {journal} {\bibinfo
   {journal} {Phys. Rev. A}\ }\textbf {\bibinfo {volume} {109}},\ \bibinfo
  {pages} {032219} (\bibinfo {year} {2024}{\natexlab{a}})}\BibitemShut
  {NoStop}%
\bibitem [{\citenamefont {Liu}\ \emph {et~al.}(2022)\citenamefont {Liu},
  \citenamefont {Ebler},\ and\ \citenamefont {Dahlsten}}]{liu}%
  \BibitemOpen
  \bibfield  {author} {\bibinfo {author} {\bibfnamefont {X.}~\bibnamefont
  {Liu}}, \bibinfo {author} {\bibfnamefont {D.}~\bibnamefont {Ebler}},\ and\
  \bibinfo {author} {\bibfnamefont {O.}~\bibnamefont {Dahlsten}},\ }\href
  {https://doi.org/10.1103/PhysRevLett.129.230604} {\bibfield  {journal}
  {\bibinfo  {journal} {Phys. Rev. Lett.}\ }\textbf {\bibinfo {volume} {129}},\
  \bibinfo {pages} {230604} (\bibinfo {year} {2022})}\BibitemShut {NoStop}%
\bibitem [{\citenamefont {Lostaglio}\ \emph
  {et~al.}(2015{\natexlab{a}})\citenamefont {Lostaglio}, \citenamefont
  {Jennings},\ and\ \citenamefont {Rudolph}}]{lostaglio2015description}%
  \BibitemOpen
  \bibfield  {author} {\bibinfo {author} {\bibfnamefont {M.}~\bibnamefont
  {Lostaglio}}, \bibinfo {author} {\bibfnamefont {D.}~\bibnamefont
  {Jennings}},\ and\ \bibinfo {author} {\bibfnamefont {T.}~\bibnamefont
  {Rudolph}},\ }\href {https://doi.org/10.1038/ncomms7383} {\bibfield
  {journal} {\bibinfo  {journal} {Nat. Commun.}\ }\textbf {\bibinfo {volume}
  {6}},\ \bibinfo {pages} {6383} (\bibinfo {year}
  {2015}{\natexlab{a}})}\BibitemShut {NoStop}%
\bibitem [{\citenamefont {Chitambar}\ and\ \citenamefont
  {Gour}(2019)}]{RevModPhys.91.025001}%
  \BibitemOpen
  \bibfield  {author} {\bibinfo {author} {\bibfnamefont {E.}~\bibnamefont
  {Chitambar}}\ and\ \bibinfo {author} {\bibfnamefont {G.}~\bibnamefont
  {Gour}},\ }\href {https://doi.org/10.1103/RevModPhys.91.025001} {\bibfield
  {journal} {\bibinfo  {journal} {Rev. Mod. Phys.}\ }\textbf {\bibinfo {volume}
  {91}},\ \bibinfo {pages} {025001} (\bibinfo {year} {2019})}\BibitemShut
  {NoStop}%
\bibitem [{\citenamefont {Brand\~ao}\ \emph {et~al.}(2013)\citenamefont
  {Brand\~ao}, \citenamefont {Horodecki}, \citenamefont {Oppenheim},
  \citenamefont {Renes},\ and\ \citenamefont
  {Spekkens}}]{Fernando2013resource}%
  \BibitemOpen
  \bibfield  {author} {\bibinfo {author} {\bibfnamefont {F.~G. S.~L.}\
  \bibnamefont {Brand\~ao}}, \bibinfo {author} {\bibfnamefont {M.}~\bibnamefont
  {Horodecki}}, \bibinfo {author} {\bibfnamefont {J.}~\bibnamefont
  {Oppenheim}}, \bibinfo {author} {\bibfnamefont {J.~M.}\ \bibnamefont
  {Renes}},\ and\ \bibinfo {author} {\bibfnamefont {R.~W.}\ \bibnamefont
  {Spekkens}},\ }\href {https://doi.org/10.1103/PhysRevLett.111.250404}
  {\bibfield  {journal} {\bibinfo  {journal} {Phys. Rev. Lett.}\ }\textbf
  {\bibinfo {volume} {111}},\ \bibinfo {pages} {250404} (\bibinfo {year}
  {2013})}\BibitemShut {NoStop}%
\bibitem [{\citenamefont {Maruyama}\ \emph {et~al.}(2009)\citenamefont
  {Maruyama}, \citenamefont {Nori},\ and\ \citenamefont
  {Vedral}}]{maruyama2009physics}%
  \BibitemOpen
  \bibfield  {author} {\bibinfo {author} {\bibfnamefont {K.}~\bibnamefont
  {Maruyama}}, \bibinfo {author} {\bibfnamefont {F.}~\bibnamefont {Nori}},\
  and\ \bibinfo {author} {\bibfnamefont {V.}~\bibnamefont {Vedral}},\ }\href
  {https://doi.org/10.1103/RevModPhys.81.1} {\bibfield  {journal} {\bibinfo
  {journal} {Rev. Mod. Phys.}\ }\textbf {\bibinfo {volume} {81}},\ \bibinfo
  {pages} {1} (\bibinfo {year} {2009})}\BibitemShut {NoStop}%
\bibitem [{\citenamefont {Bennett}(1982)}]{bennett1982thermodynamics}%
  \BibitemOpen
  \bibfield  {author} {\bibinfo {author} {\bibfnamefont {C.~H.}\ \bibnamefont
  {Bennett}},\ }\href {https://doi.org/10.1007/BF02084158} {\bibfield
  {journal} {\bibinfo  {journal} {Int. J. Theor. Phys.}\ }\textbf {\bibinfo
  {volume} {21}},\ \bibinfo {pages} {905} (\bibinfo {year} {1982})}\BibitemShut
  {NoStop}%
\bibitem [{\citenamefont {Szilard}(1929)}]{szilard1929entropieverminderung}%
  \BibitemOpen
  \bibfield  {author} {\bibinfo {author} {\bibfnamefont {L.}~\bibnamefont
  {Szilard}},\ }\href {https://doi.org/10.1007/BF01341281} {\bibfield
  {journal} {\bibinfo  {journal} {Z. Phys.}\ }\textbf {\bibinfo {volume}
  {53}},\ \bibinfo {pages} {840} (\bibinfo {year} {1929})}\BibitemShut
  {NoStop}%
\bibitem [{\citenamefont {Ji}\ \emph {et~al.}(2022)\citenamefont {Ji},
  \citenamefont {Chai}, \citenamefont {Wang}, \citenamefont {Guo},
  \citenamefont {Rong}, \citenamefont {Shi}, \citenamefont {Ren}, \citenamefont
  {Wang},\ and\ \citenamefont {Du}}]{PhysRevLett.128.090602}%
  \BibitemOpen
  \bibfield  {author} {\bibinfo {author} {\bibfnamefont {W.}~\bibnamefont
  {Ji}}, \bibinfo {author} {\bibfnamefont {Z.}~\bibnamefont {Chai}}, \bibinfo
  {author} {\bibfnamefont {M.}~\bibnamefont {Wang}}, \bibinfo {author}
  {\bibfnamefont {Y.}~\bibnamefont {Guo}}, \bibinfo {author} {\bibfnamefont
  {X.}~\bibnamefont {Rong}}, \bibinfo {author} {\bibfnamefont {F.}~\bibnamefont
  {Shi}}, \bibinfo {author} {\bibfnamefont {C.}~\bibnamefont {Ren}}, \bibinfo
  {author} {\bibfnamefont {Y.}~\bibnamefont {Wang}},\ and\ \bibinfo {author}
  {\bibfnamefont {J.}~\bibnamefont {Du}},\ }\href
  {https://doi.org/10.1103/PhysRevLett.128.090602} {\bibfield  {journal}
  {\bibinfo  {journal} {Phys. Rev. Lett.}\ }\textbf {\bibinfo {volume} {128}},\
  \bibinfo {pages} {090602} (\bibinfo {year} {2022})}\BibitemShut {NoStop}%
\bibitem [{\citenamefont {Scully}\ \emph {et~al.}(2003)\citenamefont {Scully},
  \citenamefont {Zubairy}, \citenamefont {Agarwal},\ and\ \citenamefont
  {Walther}}]{doi:10.1126/science.1078955}%
  \BibitemOpen
  \bibfield  {author} {\bibinfo {author} {\bibfnamefont {M.~O.}\ \bibnamefont
  {Scully}}, \bibinfo {author} {\bibfnamefont {M.~S.}\ \bibnamefont {Zubairy}},
  \bibinfo {author} {\bibfnamefont {G.~S.}\ \bibnamefont {Agarwal}},\ and\
  \bibinfo {author} {\bibfnamefont {H.}~\bibnamefont {Walther}},\ }\href
  {https://doi.org/10.1126/science.1078955} {\bibfield  {journal} {\bibinfo
  {journal} {Science}\ }\textbf {\bibinfo {volume} {299}},\ \bibinfo {pages}
  {862} (\bibinfo {year} {2003})}\BibitemShut {NoStop}%
\bibitem [{\citenamefont {Parrondo}\ \emph {et~al.}(2015)\citenamefont
  {Parrondo}, \citenamefont {Horowitz},\ and\ \citenamefont
  {Sagawa}}]{Parrondo2015}%
  \BibitemOpen
  \bibfield  {author} {\bibinfo {author} {\bibfnamefont {J.~M.~R.}\
  \bibnamefont {Parrondo}}, \bibinfo {author} {\bibfnamefont {J.~M.}\
  \bibnamefont {Horowitz}},\ and\ \bibinfo {author} {\bibfnamefont
  {T.}~\bibnamefont {Sagawa}},\ }\href {https://doi.org/10.1038/nphys3230}
  {\bibfield  {journal} {\bibinfo  {journal} {Nat. Phys.}\ }\textbf {\bibinfo
  {volume} {11}},\ \bibinfo {pages} {131} (\bibinfo {year} {2015})}\BibitemShut
  {NoStop}%
\bibitem [{\citenamefont {Chiribella}\ \emph {et~al.}(2013)\citenamefont
  {Chiribella}, \citenamefont {D'Ariano}, \citenamefont {Perinotti},\ and\
  \citenamefont {Valiron}}]{PhysRevA.88.022318}%
  \BibitemOpen
  \bibfield  {author} {\bibinfo {author} {\bibfnamefont {G.}~\bibnamefont
  {Chiribella}}, \bibinfo {author} {\bibfnamefont {G.~M.}\ \bibnamefont
  {D'Ariano}}, \bibinfo {author} {\bibfnamefont {P.}~\bibnamefont
  {Perinotti}},\ and\ \bibinfo {author} {\bibfnamefont {B.}~\bibnamefont
  {Valiron}},\ }\href {https://doi.org/10.1103/PhysRevA.88.022318} {\bibfield
  {journal} {\bibinfo  {journal} {Phys. Rev. A}\ }\textbf {\bibinfo {volume}
  {88}},\ \bibinfo {pages} {022318} (\bibinfo {year} {2013})}\BibitemShut
  {NoStop}%
\bibitem [{\citenamefont {Goswami}\ \emph {et~al.}(2018)\citenamefont
  {Goswami}, \citenamefont {Giarmatzi}, \citenamefont {Kewming}, \citenamefont
  {Costa}, \citenamefont {Branciard}, \citenamefont {Romero},\ and\
  \citenamefont {White}}]{PhysRevLett.121.090503}%
  \BibitemOpen
  \bibfield  {author} {\bibinfo {author} {\bibfnamefont {K.}~\bibnamefont
  {Goswami}}, \bibinfo {author} {\bibfnamefont {C.}~\bibnamefont {Giarmatzi}},
  \bibinfo {author} {\bibfnamefont {M.}~\bibnamefont {Kewming}}, \bibinfo
  {author} {\bibfnamefont {F.}~\bibnamefont {Costa}}, \bibinfo {author}
  {\bibfnamefont {C.}~\bibnamefont {Branciard}}, \bibinfo {author}
  {\bibfnamefont {J.}~\bibnamefont {Romero}},\ and\ \bibinfo {author}
  {\bibfnamefont {A.~G.}\ \bibnamefont {White}},\ }\href
  {https://doi.org/10.1103/PhysRevLett.121.090503} {\bibfield  {journal}
  {\bibinfo  {journal} {Phys. Rev. Lett.}\ }\textbf {\bibinfo {volume} {121}},\
  \bibinfo {pages} {090503} (\bibinfo {year} {2018})}\BibitemShut {NoStop}%
\bibitem [{\citenamefont {Rubino}\ \emph {et~al.}(2017)\citenamefont {Rubino},
  \citenamefont {Rozema}, \citenamefont {Feix}, \citenamefont {Ara{\'u}jo},
  \citenamefont {Zeuner}, \citenamefont {Procopio}, \citenamefont {Brukner},\
  and\ \citenamefont {Walther}}]{doi:10.1126/sciadv.1602589}%
  \BibitemOpen
  \bibfield  {author} {\bibinfo {author} {\bibfnamefont {G.}~\bibnamefont
  {Rubino}}, \bibinfo {author} {\bibfnamefont {L.~A.}\ \bibnamefont {Rozema}},
  \bibinfo {author} {\bibfnamefont {A.}~\bibnamefont {Feix}}, \bibinfo {author}
  {\bibfnamefont {M.}~\bibnamefont {Ara{\'u}jo}}, \bibinfo {author}
  {\bibfnamefont {J.~M.}\ \bibnamefont {Zeuner}}, \bibinfo {author}
  {\bibfnamefont {L.~M.}\ \bibnamefont {Procopio}}, \bibinfo {author}
  {\bibfnamefont {{\v C}.}~\bibnamefont {Brukner}},\ and\ \bibinfo {author}
  {\bibfnamefont {P.}~\bibnamefont {Walther}},\ }\href
  {https://doi.org/10.1126/sciadv.1602589} {\bibfield  {journal} {\bibinfo
  {journal} {Sci. Adv.}\ }\textbf {\bibinfo {volume} {3}},\ \bibinfo {pages}
  {e1602589} (\bibinfo {year} {2017})}\BibitemShut {NoStop}%
\bibitem [{\citenamefont {Loizeau}\ and\ \citenamefont
  {Grinbaum}(2020)}]{PhysRevA.101.012340}%
  \BibitemOpen
  \bibfield  {author} {\bibinfo {author} {\bibfnamefont {N.}~\bibnamefont
  {Loizeau}}\ and\ \bibinfo {author} {\bibfnamefont {A.}~\bibnamefont
  {Grinbaum}},\ }\href {https://doi.org/10.1103/PhysRevA.101.012340} {\bibfield
   {journal} {\bibinfo  {journal} {Phys. Rev. A}\ }\textbf {\bibinfo {volume}
  {101}},\ \bibinfo {pages} {012340} (\bibinfo {year} {2020})}\BibitemShut
  {NoStop}%
\bibitem [{\citenamefont {Gu{\'e}rin}\ \emph {et~al.}(2016)\citenamefont
  {Gu{\'e}rin}, \citenamefont {Feix}, \citenamefont {Ara{\'u}jo},\ and\
  \citenamefont {Brukner}}]{guerin2016exponential}%
  \BibitemOpen
  \bibfield  {author} {\bibinfo {author} {\bibfnamefont {P.~A.}\ \bibnamefont
  {Gu{\'e}rin}}, \bibinfo {author} {\bibfnamefont {A.}~\bibnamefont {Feix}},
  \bibinfo {author} {\bibfnamefont {M.}~\bibnamefont {Ara{\'u}jo}},\ and\
  \bibinfo {author} {\bibfnamefont {{\v{C}}.}~\bibnamefont {Brukner}},\ }\href
  {https://link.aps.org/doi/10.1103/PhysRevLett.117.100502} {\bibfield
  {journal} {\bibinfo  {journal} {Phys. Rev. Lett}\ }\textbf {\bibinfo {volume}
  {117}},\ \bibinfo {pages} {100502} (\bibinfo {year} {2016})}\BibitemShut
  {NoStop}%
\bibitem [{\citenamefont {Renner}\ and\ \citenamefont
  {Brukner}(2022)}]{renner2022computational}%
  \BibitemOpen
  \bibfield  {author} {\bibinfo {author} {\bibfnamefont {M.~J.}\ \bibnamefont
  {Renner}}\ and\ \bibinfo {author} {\bibfnamefont {{\v{C}}.}~\bibnamefont
  {Brukner}},\ }\href {https://doi.org/10.1103/PhysRevLett.128.230503}
  {\bibfield  {journal} {\bibinfo  {journal} {Phys. Rev. Lett.}\ }\textbf
  {\bibinfo {volume} {128}},\ \bibinfo {pages} {230503} (\bibinfo {year}
  {2022})}\BibitemShut {NoStop}%
\bibitem [{\citenamefont {Ebler}\ \emph {et~al.}(2018)\citenamefont {Ebler},
  \citenamefont {Salek},\ and\ \citenamefont {Chiribella}}]{ebler2018enhance}%
  \BibitemOpen
  \bibfield  {author} {\bibinfo {author} {\bibfnamefont {D.}~\bibnamefont
  {Ebler}}, \bibinfo {author} {\bibfnamefont {S.}~\bibnamefont {Salek}},\ and\
  \bibinfo {author} {\bibfnamefont {G.}~\bibnamefont {Chiribella}},\ }\href
  {https://doi.org/10.1103/PhysRevLett.120.120502} {\bibfield  {journal}
  {\bibinfo  {journal} {Phys. Rev. Lett.}\ }\textbf {\bibinfo {volume} {120}},\
  \bibinfo {pages} {120502} (\bibinfo {year} {2018})}\BibitemShut {NoStop}%
\bibitem [{\citenamefont {Zhao}\ \emph
  {et~al.}(2020{\natexlab{a}})\citenamefont {Zhao}, \citenamefont {Yang},\ and\
  \citenamefont {Chiribella}}]{zhao2020quantum}%
  \BibitemOpen
  \bibfield  {author} {\bibinfo {author} {\bibfnamefont {X.}~\bibnamefont
  {Zhao}}, \bibinfo {author} {\bibfnamefont {Y.}~\bibnamefont {Yang}},\ and\
  \bibinfo {author} {\bibfnamefont {G.}~\bibnamefont {Chiribella}},\ }\href
  {https://doi.org/10.1103/PhysRevLett.124.190503} {\bibfield  {journal}
  {\bibinfo  {journal} {Phys. Rev. Lett.}\ }\textbf {\bibinfo {volume} {124}},\
  \bibinfo {pages} {190503} (\bibinfo {year} {2020}{\natexlab{a}})}\BibitemShut
  {NoStop}%
\bibitem [{\citenamefont {Yin}\ \emph {et~al.}(2023)\citenamefont {Yin},
  \citenamefont {Zhao}, \citenamefont {Yang}, \citenamefont {Guo},
  \citenamefont {Zhang}, \citenamefont {Li}, \citenamefont {Han}, \citenamefont
  {Liu}, \citenamefont {Xu}, \citenamefont {Chiribella} \emph
  {et~al.}}]{yin2023experimental}%
  \BibitemOpen
  \bibfield  {author} {\bibinfo {author} {\bibfnamefont {P.}~\bibnamefont
  {Yin}}, \bibinfo {author} {\bibfnamefont {X.}~\bibnamefont {Zhao}}, \bibinfo
  {author} {\bibfnamefont {Y.}~\bibnamefont {Yang}}, \bibinfo {author}
  {\bibfnamefont {Y.}~\bibnamefont {Guo}}, \bibinfo {author} {\bibfnamefont
  {W.-H.}\ \bibnamefont {Zhang}}, \bibinfo {author} {\bibfnamefont {G.-C.}\
  \bibnamefont {Li}}, \bibinfo {author} {\bibfnamefont {Y.-J.}\ \bibnamefont
  {Han}}, \bibinfo {author} {\bibfnamefont {B.-H.}\ \bibnamefont {Liu}},
  \bibinfo {author} {\bibfnamefont {J.-S.}\ \bibnamefont {Xu}}, \bibinfo
  {author} {\bibfnamefont {G.}~\bibnamefont {Chiribella}}, \emph {et~al.},\
  }\href {https://doi.org/10.1038/s41567-023-02046-y} {\bibfield  {journal}
  {\bibinfo  {journal} {Nat. Phys.}\ ,\ \bibinfo {pages} {1}} (\bibinfo {year}
  {2023})}\BibitemShut {NoStop}%
\bibitem [{\citenamefont {Liu}\ \emph {et~al.}(2023)\citenamefont {Liu},
  \citenamefont {Meng}, \citenamefont {Song}, \citenamefont {Li}, \citenamefont
  {Wu}, \citenamefont {Chen}, \citenamefont {Hong}, \citenamefont {Zhang},\
  and\ \citenamefont {Yin}}]{liu2023experimentally}%
  \BibitemOpen
  \bibfield  {author} {\bibinfo {author} {\bibfnamefont {W.-Q.}\ \bibnamefont
  {Liu}}, \bibinfo {author} {\bibfnamefont {Z.}~\bibnamefont {Meng}}, \bibinfo
  {author} {\bibfnamefont {B.-W.}\ \bibnamefont {Song}}, \bibinfo {author}
  {\bibfnamefont {J.}~\bibnamefont {Li}}, \bibinfo {author} {\bibfnamefont
  {Q.-Y.}\ \bibnamefont {Wu}}, \bibinfo {author} {\bibfnamefont {X.-X.}\
  \bibnamefont {Chen}}, \bibinfo {author} {\bibfnamefont {J.-Y.}\ \bibnamefont
  {Hong}}, \bibinfo {author} {\bibfnamefont {A.-N.}\ \bibnamefont {Zhang}},\
  and\ \bibinfo {author} {\bibfnamefont {Z.-Q.}\ \bibnamefont {Yin}},\ }\href
  {https://doi.org/10.48550/arXiv.2305.05416} {\bibfield  {journal} {\bibinfo
  {journal} {arXiv:2305.05416}\ } (\bibinfo {year} {2023})}\BibitemShut
  {NoStop}%
\bibitem [{\citenamefont {Liu}\ \emph {et~al.}(2024{\natexlab{b}})\citenamefont
  {Liu}, \citenamefont {Jia}, \citenamefont {Qiu}, \citenamefont {Li},\ and\
  \citenamefont {Dahlsten}}]{liu2023unification}%
  \BibitemOpen
  \bibfield  {author} {\bibinfo {author} {\bibfnamefont {X.}~\bibnamefont
  {Liu}}, \bibinfo {author} {\bibfnamefont {Z.}~\bibnamefont {Jia}}, \bibinfo
  {author} {\bibfnamefont {Y.}~\bibnamefont {Qiu}}, \bibinfo {author}
  {\bibfnamefont {F.}~\bibnamefont {Li}},\ and\ \bibinfo {author}
  {\bibfnamefont {O.}~\bibnamefont {Dahlsten}},\ }\href
  {https://iopscience.iop.org/article/10.1088/1367-2630/ad264c} {\bibfield
  {journal} {\bibinfo  {journal} {New Journal of Physics}\ }\textbf {\bibinfo
  {volume} {26}},\ \bibinfo {pages} {033008} (\bibinfo {year}
  {2024}{\natexlab{b}})}\BibitemShut {NoStop}%
\bibitem [{\citenamefont {Felce}\ and\ \citenamefont
  {Vedral}(2020{\natexlab{a}})}]{PhysRevLett.125.070603}%
  \BibitemOpen
  \bibfield  {author} {\bibinfo {author} {\bibfnamefont {D.}~\bibnamefont
  {Felce}}\ and\ \bibinfo {author} {\bibfnamefont {V.}~\bibnamefont {Vedral}},\
  }\href {https://doi.org/10.1103/PhysRevLett.125.070603} {\bibfield  {journal}
  {\bibinfo  {journal} {Phys. Rev. Lett.}\ }\textbf {\bibinfo {volume} {125}},\
  \bibinfo {pages} {070603} (\bibinfo {year} {2020}{\natexlab{a}})}\BibitemShut
  {NoStop}%
\bibitem [{\citenamefont {Zhu}\ \emph {et~al.}(2023)\citenamefont {Zhu},
  \citenamefont {Chen}, \citenamefont {Hasegawa},\ and\ \citenamefont
  {Xue}}]{zhu2023charging}%
  \BibitemOpen
  \bibfield  {author} {\bibinfo {author} {\bibfnamefont {G.}~\bibnamefont
  {Zhu}}, \bibinfo {author} {\bibfnamefont {Y.}~\bibnamefont {Chen}}, \bibinfo
  {author} {\bibfnamefont {Y.}~\bibnamefont {Hasegawa}},\ and\ \bibinfo
  {author} {\bibfnamefont {P.}~\bibnamefont {Xue}},\ }\href
  {https://doi.org/10.1103/PhysRevLett.131.240401} {\bibfield  {journal}
  {\bibinfo  {journal} {Phys. Rev. Lett.}\ }\textbf {\bibinfo {volume} {131}},\
  \bibinfo {pages} {240401} (\bibinfo {year} {2023})}\BibitemShut {NoStop}%
\bibitem [{\citenamefont {Cover}(1999)}]{cover1999elements}%
  \BibitemOpen
  \bibfield  {author} {\bibinfo {author} {\bibfnamefont {T.~M.}\ \bibnamefont
  {Cover}},\ }\href@noop {} {\emph {\bibinfo {title} {Elements of information
  theory}}}\ (\bibinfo  {publisher} {John Wiley \& Sons},\ \bibinfo {year}
  {1999})\BibitemShut {NoStop}%
\bibitem [{\citenamefont {Wilde}(2013)}]{wilde2013quantum}%
  \BibitemOpen
  \bibfield  {author} {\bibinfo {author} {\bibfnamefont {M.~M.}\ \bibnamefont
  {Wilde}},\ }\href@noop {} {\emph {\bibinfo {title} {Quantum information
  theory}}}\ (\bibinfo  {publisher} {Cambridge university press},\ \bibinfo
  {year} {2013})\BibitemShut {NoStop}%
\bibitem [{\citenamefont {Salek}\ \emph {et~al.}()\citenamefont {Salek},
  \citenamefont {Ebler},\ and\ \citenamefont {Chiribella}}]{salek2018quantum}%
  \BibitemOpen
  \bibfield  {author} {\bibinfo {author} {\bibfnamefont {S.}~\bibnamefont
  {Salek}}, \bibinfo {author} {\bibfnamefont {D.}~\bibnamefont {Ebler}},\ and\
  \bibinfo {author} {\bibfnamefont {G.}~\bibnamefont {Chiribella}},\
  }\href@noop {} {}\Eprint {https://arxiv.org/abs/1809.06655}
  {arXiv:1809.06655} \BibitemShut {NoStop}%
\bibitem [{\citenamefont {Caleffi}\ and\ \citenamefont
  {Cacciapuoti}(2020)}]{8966996}%
  \BibitemOpen
  \bibfield  {author} {\bibinfo {author} {\bibfnamefont {M.}~\bibnamefont
  {Caleffi}}\ and\ \bibinfo {author} {\bibfnamefont {A.~S.}\ \bibnamefont
  {Cacciapuoti}},\ }\href {https://doi.org/10.1109/JSAC.2020.2969035}
  {\bibfield  {journal} {\bibinfo  {journal} {IEEE J. Sel. Areas Commun.}\
  }\textbf {\bibinfo {volume} {38}},\ \bibinfo {pages} {575} (\bibinfo {year}
  {2020})}\BibitemShut {NoStop}%
\bibitem [{\citenamefont {Simonov}\ \emph {et~al.}(2022)\citenamefont
  {Simonov}, \citenamefont {Francica}, \citenamefont {Guarnieri},\ and\
  \citenamefont {Paternostro}}]{Simonov2022work}%
  \BibitemOpen
  \bibfield  {author} {\bibinfo {author} {\bibfnamefont {K.}~\bibnamefont
  {Simonov}}, \bibinfo {author} {\bibfnamefont {G.}~\bibnamefont {Francica}},
  \bibinfo {author} {\bibfnamefont {G.}~\bibnamefont {Guarnieri}},\ and\
  \bibinfo {author} {\bibfnamefont {M.}~\bibnamefont {Paternostro}},\ }\href
  {https://doi.org/10.1103/PhysRevA.105.032217} {\bibfield  {journal} {\bibinfo
   {journal} {Phys. Rev. A}\ }\textbf {\bibinfo {volume} {105}},\ \bibinfo
  {pages} {032217} (\bibinfo {year} {2022})}\BibitemShut {NoStop}%
\bibitem [{\citenamefont {Verma}\ \emph {et~al.}(2024)\citenamefont {Verma},
  \citenamefont {Zych},\ and\ \citenamefont {Costa}}]{verma2024measuring}%
  \BibitemOpen
  \bibfield  {author} {\bibinfo {author} {\bibfnamefont {H.}~\bibnamefont
  {Verma}}, \bibinfo {author} {\bibfnamefont {M.}~\bibnamefont {Zych}},\ and\
  \bibinfo {author} {\bibfnamefont {F.}~\bibnamefont {Costa}},\ }\href
  {https://doi.org/10.48550/arXiv.2403.15186} {\bibfield  {journal} {\bibinfo
  {journal} {arXiv:2403.15186}\ } (\bibinfo {year} {2024})}\BibitemShut
  {NoStop}%
\bibitem [{\citenamefont {Dieguez}\ \emph {et~al.}(2023)\citenamefont
  {Dieguez}, \citenamefont {Lisboa},\ and\ \citenamefont
  {Serra}}]{dieguez2023thermal}%
  \BibitemOpen
  \bibfield  {author} {\bibinfo {author} {\bibfnamefont {P.~R.}\ \bibnamefont
  {Dieguez}}, \bibinfo {author} {\bibfnamefont {V.~F.}\ \bibnamefont
  {Lisboa}},\ and\ \bibinfo {author} {\bibfnamefont {R.~M.}\ \bibnamefont
  {Serra}},\ }\href {https://doi.org/10.1103/PhysRevA.107.012423} {\bibfield
  {journal} {\bibinfo  {journal} {Phys. Rev. A}\ }\textbf {\bibinfo {volume}
  {107}},\ \bibinfo {pages} {012423} (\bibinfo {year} {2023})}\BibitemShut
  {NoStop}%
\bibitem [{\citenamefont {Guha}\ \emph {et~al.}(2020)\citenamefont {Guha},
  \citenamefont {Alimuddin},\ and\ \citenamefont
  {Parashar}}]{Guha2020thermodynamics}%
  \BibitemOpen
  \bibfield  {author} {\bibinfo {author} {\bibfnamefont {T.}~\bibnamefont
  {Guha}}, \bibinfo {author} {\bibfnamefont {M.}~\bibnamefont {Alimuddin}},\
  and\ \bibinfo {author} {\bibfnamefont {P.}~\bibnamefont {Parashar}},\ }\href
  {https://doi.org/10.1103/PhysRevA.102.032215} {\bibfield  {journal} {\bibinfo
   {journal} {Phys. Rev. A}\ }\textbf {\bibinfo {volume} {102}},\ \bibinfo
  {pages} {032215} (\bibinfo {year} {2020})}\BibitemShut {NoStop}%
\bibitem [{\citenamefont {Capela}\ \emph {et~al.}(2023)\citenamefont {Capela},
  \citenamefont {Verma}, \citenamefont {Costa},\ and\ \citenamefont
  {C\'eleri}}]{capela2023reassessing}%
  \BibitemOpen
  \bibfield  {author} {\bibinfo {author} {\bibfnamefont {M.}~\bibnamefont
  {Capela}}, \bibinfo {author} {\bibfnamefont {H.}~\bibnamefont {Verma}},
  \bibinfo {author} {\bibfnamefont {F.}~\bibnamefont {Costa}},\ and\ \bibinfo
  {author} {\bibfnamefont {L.~C.}\ \bibnamefont {C\'eleri}},\ }\href
  {https://doi.org/10.1103/PhysRevA.107.062208} {\bibfield  {journal} {\bibinfo
   {journal} {Phys. Rev. A}\ }\textbf {\bibinfo {volume} {107}},\ \bibinfo
  {pages} {062208} (\bibinfo {year} {2023})}\BibitemShut {NoStop}%
\bibitem [{\citenamefont {Felce}\ and\ \citenamefont
  {Vedral}(2020{\natexlab{b}})}]{felce2020quantum}%
  \BibitemOpen
  \bibfield  {author} {\bibinfo {author} {\bibfnamefont {D.}~\bibnamefont
  {Felce}}\ and\ \bibinfo {author} {\bibfnamefont {V.}~\bibnamefont {Vedral}},\
  }\href {https://doi.org/10.1103/PhysRevLett.125.070603} {\bibfield  {journal}
  {\bibinfo  {journal} {Phys. Rev. Lett.}\ }\textbf {\bibinfo {volume} {125}},\
  \bibinfo {pages} {070603} (\bibinfo {year} {2020}{\natexlab{b}})}\BibitemShut
  {NoStop}%
\bibitem [{\citenamefont {Nie}\ \emph {et~al.}(2022)\citenamefont {Nie},
  \citenamefont {Zhu}, \citenamefont {Huang}, \citenamefont {Tang},
  \citenamefont {Long}, \citenamefont {Lin}, \citenamefont {Tian},
  \citenamefont {Qiu}, \citenamefont {Xi}, \citenamefont {Yang}, \citenamefont
  {Li}, \citenamefont {Dong}, \citenamefont {Xin},\ and\ \citenamefont
  {Lu}}]{PhysRevLett.129.100603}%
  \BibitemOpen
  \bibfield  {author} {\bibinfo {author} {\bibfnamefont {X.}~\bibnamefont
  {Nie}}, \bibinfo {author} {\bibfnamefont {X.}~\bibnamefont {Zhu}}, \bibinfo
  {author} {\bibfnamefont {K.}~\bibnamefont {Huang}}, \bibinfo {author}
  {\bibfnamefont {K.}~\bibnamefont {Tang}}, \bibinfo {author} {\bibfnamefont
  {X.}~\bibnamefont {Long}}, \bibinfo {author} {\bibfnamefont {Z.}~\bibnamefont
  {Lin}}, \bibinfo {author} {\bibfnamefont {Y.}~\bibnamefont {Tian}}, \bibinfo
  {author} {\bibfnamefont {C.}~\bibnamefont {Qiu}}, \bibinfo {author}
  {\bibfnamefont {C.}~\bibnamefont {Xi}}, \bibinfo {author} {\bibfnamefont
  {X.}~\bibnamefont {Yang}}, \bibinfo {author} {\bibfnamefont {J.}~\bibnamefont
  {Li}}, \bibinfo {author} {\bibfnamefont {Y.}~\bibnamefont {Dong}}, \bibinfo
  {author} {\bibfnamefont {T.}~\bibnamefont {Xin}},\ and\ \bibinfo {author}
  {\bibfnamefont {D.}~\bibnamefont {Lu}},\ }\href
  {https://doi.org/10.1103/PhysRevLett.129.100603} {\bibfield  {journal}
  {\bibinfo  {journal} {Phys. Rev. Lett.}\ }\textbf {\bibinfo {volume} {129}},\
  \bibinfo {pages} {100603} (\bibinfo {year} {2022})}\BibitemShut {NoStop}%
\bibitem [{\citenamefont {Cao}\ \emph {et~al.}(2022)\citenamefont {Cao},
  \citenamefont {Wang}, \citenamefont {Jia}, \citenamefont {Zhang},
  \citenamefont {Guo}, \citenamefont {Liu}, \citenamefont {Huang},
  \citenamefont {Li},\ and\ \citenamefont {Guo}}]{cao2022quantum}%
  \BibitemOpen
  \bibfield  {author} {\bibinfo {author} {\bibfnamefont {H.}~\bibnamefont
  {Cao}}, \bibinfo {author} {\bibfnamefont {N.-N.}\ \bibnamefont {Wang}},
  \bibinfo {author} {\bibfnamefont {Z.}~\bibnamefont {Jia}}, \bibinfo {author}
  {\bibfnamefont {C.}~\bibnamefont {Zhang}}, \bibinfo {author} {\bibfnamefont
  {Y.}~\bibnamefont {Guo}}, \bibinfo {author} {\bibfnamefont {B.-H.}\
  \bibnamefont {Liu}}, \bibinfo {author} {\bibfnamefont {Y.-F.}\ \bibnamefont
  {Huang}}, \bibinfo {author} {\bibfnamefont {C.-F.}\ \bibnamefont {Li}},\ and\
  \bibinfo {author} {\bibfnamefont {G.-C.}\ \bibnamefont {Guo}},\ }\href
  {https://doi.org/10.1103/PhysRevResearch.4.L032029} {\bibfield  {journal}
  {\bibinfo  {journal} {Phys. Rev. Res.}\ }\textbf {\bibinfo {volume} {4}},\
  \bibinfo {pages} {L032029} (\bibinfo {year} {2022})}\BibitemShut {NoStop}%
\bibitem [{\citenamefont {Micadei}\ \emph {et~al.}(2019)\citenamefont
  {Micadei}, \citenamefont {Peterson}, \citenamefont {Souza}, \citenamefont
  {Sarthour}, \citenamefont {Oliveira}, \citenamefont {Landi}, \citenamefont
  {Batalh{\~a}o}, \citenamefont {Serra},\ and\ \citenamefont
  {Lutz}}]{Micadei2019}%
  \BibitemOpen
  \bibfield  {author} {\bibinfo {author} {\bibfnamefont {K.}~\bibnamefont
  {Micadei}}, \bibinfo {author} {\bibfnamefont {J.~P.~S.}\ \bibnamefont
  {Peterson}}, \bibinfo {author} {\bibfnamefont {A.~M.}\ \bibnamefont {Souza}},
  \bibinfo {author} {\bibfnamefont {R.~S.}\ \bibnamefont {Sarthour}}, \bibinfo
  {author} {\bibfnamefont {I.~S.}\ \bibnamefont {Oliveira}}, \bibinfo {author}
  {\bibfnamefont {G.~T.}\ \bibnamefont {Landi}}, \bibinfo {author}
  {\bibfnamefont {T.~B.}\ \bibnamefont {Batalh{\~a}o}}, \bibinfo {author}
  {\bibfnamefont {R.~M.}\ \bibnamefont {Serra}},\ and\ \bibinfo {author}
  {\bibfnamefont {E.}~\bibnamefont {Lutz}},\ }\href
  {https://doi.org/10.1038/s41467-019-10333-7} {\bibfield  {journal} {\bibinfo
  {journal} {Nat. Commun.}\ }\textbf {\bibinfo {volume} {10}},\ \bibinfo
  {pages} {2456} (\bibinfo {year} {2019})}\BibitemShut {NoStop}%
\bibitem [{\citenamefont {Peterson}\ \emph {et~al.}(2019)\citenamefont
  {Peterson}, \citenamefont {Batalh\~ao}, \citenamefont {Herrera},
  \citenamefont {Souza}, \citenamefont {Sarthour}, \citenamefont {Oliveira},\
  and\ \citenamefont {Serra}}]{PhysRevLett.123.240601}%
  \BibitemOpen
  \bibfield  {author} {\bibinfo {author} {\bibfnamefont {J.~P.~S.}\
  \bibnamefont {Peterson}}, \bibinfo {author} {\bibfnamefont {T.~B.}\
  \bibnamefont {Batalh\~ao}}, \bibinfo {author} {\bibfnamefont
  {M.}~\bibnamefont {Herrera}}, \bibinfo {author} {\bibfnamefont {A.~M.}\
  \bibnamefont {Souza}}, \bibinfo {author} {\bibfnamefont {R.~S.}\ \bibnamefont
  {Sarthour}}, \bibinfo {author} {\bibfnamefont {I.~S.}\ \bibnamefont
  {Oliveira}},\ and\ \bibinfo {author} {\bibfnamefont {R.~M.}\ \bibnamefont
  {Serra}},\ }\href {https://doi.org/10.1103/PhysRevLett.123.240601} {\bibfield
   {journal} {\bibinfo  {journal} {Phys. Rev. Lett.}\ }\textbf {\bibinfo
  {volume} {123}},\ \bibinfo {pages} {240601} (\bibinfo {year}
  {2019})}\BibitemShut {NoStop}%
\bibitem [{\citenamefont {Vieira}\ \emph {et~al.}(2023)\citenamefont {Vieira},
  \citenamefont {{de Oliveira}}, \citenamefont {Santos}, \citenamefont
  {Dieguez},\ and\ \citenamefont {Serra}}]{VIEIRA2023100105}%
  \BibitemOpen
  \bibfield  {author} {\bibinfo {author} {\bibfnamefont {C.}~\bibnamefont
  {Vieira}}, \bibinfo {author} {\bibfnamefont {J.}~\bibnamefont {{de
  Oliveira}}}, \bibinfo {author} {\bibfnamefont {J.}~\bibnamefont {Santos}},
  \bibinfo {author} {\bibfnamefont {P.}~\bibnamefont {Dieguez}},\ and\ \bibinfo
  {author} {\bibfnamefont {R.}~\bibnamefont {Serra}},\ }\href
  {https://doi.org/https://doi.org/10.1016/j.jmro.2023.100105} {\bibfield
  {journal} {\bibinfo  {journal} {J. Magn. Reson. Open}\ }\textbf {\bibinfo
  {volume} {16-17}},\ \bibinfo {pages} {100105} (\bibinfo {year}
  {2023})}\BibitemShut {NoStop}%
\bibitem [{\citenamefont {Scarani}\ \emph {et~al.}(2002)\citenamefont
  {Scarani}, \citenamefont {Ziman}, \citenamefont {\ifmmode \check{S}\else
  \v{S}\fi{}telmachovi\ifmmode~\check{c}\else \v{c}\fi{}}, \citenamefont
  {Gisin},\ and\ \citenamefont {Bu\ifmmode~\check{z}\else
  \v{z}\fi{}ek}}]{scarani2002thermalizing}%
  \BibitemOpen
  \bibfield  {author} {\bibinfo {author} {\bibfnamefont {V.}~\bibnamefont
  {Scarani}}, \bibinfo {author} {\bibfnamefont {M.}~\bibnamefont {Ziman}},
  \bibinfo {author} {\bibfnamefont {P.}~\bibnamefont {\ifmmode \check{S}\else
  \v{S}\fi{}telmachovi\ifmmode~\check{c}\else \v{c}\fi{}}}, \bibinfo {author}
  {\bibfnamefont {N.}~\bibnamefont {Gisin}},\ and\ \bibinfo {author}
  {\bibfnamefont {V.}~\bibnamefont {Bu\ifmmode~\check{z}\else \v{z}\fi{}ek}},\
  }\href {https://doi.org/10.1103/PhysRevLett.88.097905} {\bibfield  {journal}
  {\bibinfo  {journal} {Phys. Rev. Lett.}\ }\textbf {\bibinfo {volume} {88}},\
  \bibinfo {pages} {097905} (\bibinfo {year} {2002})}\BibitemShut {NoStop}%
\bibitem [{\citenamefont {Kraus}\ and\ \citenamefont
  {Cirac}(2001)}]{kraus2001optimal}%
  \BibitemOpen
  \bibfield  {author} {\bibinfo {author} {\bibfnamefont {B.}~\bibnamefont
  {Kraus}}\ and\ \bibinfo {author} {\bibfnamefont {J.~I.}\ \bibnamefont
  {Cirac}},\ }\href {https://doi.org/10.1103/PhysRevA.63.062309} {\bibfield
  {journal} {\bibinfo  {journal} {Phys. Rev. A}\ }\textbf {\bibinfo {volume}
  {63}},\ \bibinfo {pages} {062309} (\bibinfo {year} {2001})}\BibitemShut
  {NoStop}%
\bibitem [{sup()}]{supp}%
  \BibitemOpen
  \href@noop {} {}\bibinfo {note} {See Supplemental Materials for more details
  on theoretical analysis and experimental procedures, which includes
  additional
  Refs.~\cite{Xin_2018,RevModPhys.76.1037,JONES201191,KHANEJA2005296,park2016simulation}.}\BibitemShut
  {Stop}%
\bibitem [{\citenamefont {Xin}\ \emph {et~al.}(2018)\citenamefont {Xin},
  \citenamefont {Wang}, \citenamefont {Li}, \citenamefont {Kong}, \citenamefont
  {Wei}, \citenamefont {Wang}, \citenamefont {Ruan},\ and\ \citenamefont
  {Long}}]{Xin_2018}%
  \BibitemOpen
  \bibfield  {author} {\bibinfo {author} {\bibfnamefont {T.}~\bibnamefont
  {Xin}}, \bibinfo {author} {\bibfnamefont {B.-X.}\ \bibnamefont {Wang}},
  \bibinfo {author} {\bibfnamefont {K.-R.}\ \bibnamefont {Li}}, \bibinfo
  {author} {\bibfnamefont {X.-Y.}\ \bibnamefont {Kong}}, \bibinfo {author}
  {\bibfnamefont {S.-J.}\ \bibnamefont {Wei}}, \bibinfo {author} {\bibfnamefont
  {T.}~\bibnamefont {Wang}}, \bibinfo {author} {\bibfnamefont {D.}~\bibnamefont
  {Ruan}},\ and\ \bibinfo {author} {\bibfnamefont {G.-L.}\ \bibnamefont
  {Long}},\ }\href {https://doi.org/10.1088/1674-1056/27/2/020308} {\bibfield
  {journal} {\bibinfo  {journal} {Chin. Phys. B}\ }\textbf {\bibinfo {volume}
  {27}},\ \bibinfo {pages} {020308} (\bibinfo {year} {2018})}\BibitemShut
  {NoStop}%
\bibitem [{\citenamefont {Vandersypen}\ and\ \citenamefont
  {Chuang}(2005)}]{RevModPhys.76.1037}%
  \BibitemOpen
  \bibfield  {author} {\bibinfo {author} {\bibfnamefont {L.~M.~K.}\
  \bibnamefont {Vandersypen}}\ and\ \bibinfo {author} {\bibfnamefont {I.~L.}\
  \bibnamefont {Chuang}},\ }\href {https://doi.org/10.1103/RevModPhys.76.1037}
  {\bibfield  {journal} {\bibinfo  {journal} {Rev. Mod. Phys.}\ }\textbf
  {\bibinfo {volume} {76}},\ \bibinfo {pages} {1037} (\bibinfo {year}
  {2005})}\BibitemShut {NoStop}%
\bibitem [{\citenamefont {Jones}(2011)}]{JONES201191}%
  \BibitemOpen
  \bibfield  {author} {\bibinfo {author} {\bibfnamefont {J.~A.}\ \bibnamefont
  {Jones}},\ }\href
  {https://doi.org/https://doi.org/10.1016/j.pnmrs.2010.11.001} {\bibfield
  {journal} {\bibinfo  {journal} {Prog. Nucl. Magn. Reson. Spectrosc.}\
  }\textbf {\bibinfo {volume} {59}},\ \bibinfo {pages} {91} (\bibinfo {year}
  {2011})}\BibitemShut {NoStop}%
\bibitem [{\citenamefont {Khaneja}\ \emph {et~al.}(2005)\citenamefont
  {Khaneja}, \citenamefont {Reiss}, \citenamefont {Kehlet}, \citenamefont
  {Schulte-Herbrüggen},\ and\ \citenamefont {Glaser}}]{KHANEJA2005296}%
  \BibitemOpen
  \bibfield  {author} {\bibinfo {author} {\bibfnamefont {N.}~\bibnamefont
  {Khaneja}}, \bibinfo {author} {\bibfnamefont {T.}~\bibnamefont {Reiss}},
  \bibinfo {author} {\bibfnamefont {C.}~\bibnamefont {Kehlet}}, \bibinfo
  {author} {\bibfnamefont {T.}~\bibnamefont {Schulte-Herbrüggen}},\ and\
  \bibinfo {author} {\bibfnamefont {S.~J.}\ \bibnamefont {Glaser}},\ }\href
  {https://doi.org/https://doi.org/10.1016/j.jmr.2004.11.004} {\bibfield
  {journal} {\bibinfo  {journal} {J. Magn. Reson.}\ }\textbf {\bibinfo {volume}
  {172}},\ \bibinfo {pages} {296} (\bibinfo {year} {2005})}\BibitemShut
  {NoStop}%
\bibitem [{\citenamefont {Park}\ \emph {et~al.}(2016)\citenamefont {Park},
  \citenamefont {McKay}, \citenamefont {Lu},\ and\ \citenamefont
  {Laflamme}}]{park2016simulation}%
  \BibitemOpen
  \bibfield  {author} {\bibinfo {author} {\bibfnamefont {A.~J.}\ \bibnamefont
  {Park}}, \bibinfo {author} {\bibfnamefont {E.}~\bibnamefont {McKay}},
  \bibinfo {author} {\bibfnamefont {D.}~\bibnamefont {Lu}},\ and\ \bibinfo
  {author} {\bibfnamefont {R.}~\bibnamefont {Laflamme}},\ }\href
  {https://doi.org/10.1088/1367-2630/18/4/043043} {\bibfield  {journal}
  {\bibinfo  {journal} {New J. Phys.}\ }\textbf {\bibinfo {volume} {18}},\
  \bibinfo {pages} {043043} (\bibinfo {year} {2016})}\BibitemShut {NoStop}%
\bibitem [{\citenamefont {Chiribella}\ \emph {et~al.}(2008)\citenamefont
  {Chiribella}, \citenamefont {D'Ariano},\ and\ \citenamefont
  {Perinotti}}]{PhysRevLett.101.060401}%
  \BibitemOpen
  \bibfield  {author} {\bibinfo {author} {\bibfnamefont {G.}~\bibnamefont
  {Chiribella}}, \bibinfo {author} {\bibfnamefont {G.~M.}\ \bibnamefont
  {D'Ariano}},\ and\ \bibinfo {author} {\bibfnamefont {P.}~\bibnamefont
  {Perinotti}},\ }\href {https://doi.org/10.1103/PhysRevLett.101.060401}
  {\bibfield  {journal} {\bibinfo  {journal} {Phys. Rev. Lett.}\ }\textbf
  {\bibinfo {volume} {101}},\ \bibinfo {pages} {060401} (\bibinfo {year}
  {2008})}\BibitemShut {NoStop}%
\bibitem [{\citenamefont {Zhao}\ \emph
  {et~al.}(2020{\natexlab{b}})\citenamefont {Zhao}, \citenamefont {Yang},\ and\
  \citenamefont {Chiribella}}]{PhysRevLett.124.190503}%
  \BibitemOpen
  \bibfield  {author} {\bibinfo {author} {\bibfnamefont {X.}~\bibnamefont
  {Zhao}}, \bibinfo {author} {\bibfnamefont {Y.}~\bibnamefont {Yang}},\ and\
  \bibinfo {author} {\bibfnamefont {G.}~\bibnamefont {Chiribella}},\ }\href
  {https://doi.org/10.1103/PhysRevLett.124.190503} {\bibfield  {journal}
  {\bibinfo  {journal} {Phys. Rev. Lett.}\ }\textbf {\bibinfo {volume} {124}},\
  \bibinfo {pages} {190503} (\bibinfo {year} {2020}{\natexlab{b}})}\BibitemShut
  {NoStop}%
\bibitem [{\citenamefont {Janzing}\ \emph {et~al.}(2000)\citenamefont
  {Janzing}, \citenamefont {Wocjan}, \citenamefont {Zeier}, \citenamefont
  {Geiss},\ and\ \citenamefont {Beth}}]{janzing2000thermodynamic}%
  \BibitemOpen
  \bibfield  {author} {\bibinfo {author} {\bibfnamefont {D.}~\bibnamefont
  {Janzing}}, \bibinfo {author} {\bibfnamefont {P.}~\bibnamefont {Wocjan}},
  \bibinfo {author} {\bibfnamefont {R.}~\bibnamefont {Zeier}}, \bibinfo
  {author} {\bibfnamefont {R.}~\bibnamefont {Geiss}},\ and\ \bibinfo {author}
  {\bibfnamefont {T.}~\bibnamefont {Beth}},\ }\href
  {https://doi.org/10.1023/A:1026422630734} {\bibfield  {journal} {\bibinfo
  {journal} {Int. J. Theor. Phys.}\ }\textbf {\bibinfo {volume} {39}},\
  \bibinfo {pages} {2717} (\bibinfo {year} {2000})}\BibitemShut {NoStop}%
\bibitem [{\citenamefont {Long}\ \emph {et~al.}(2022)\citenamefont {Long},
  \citenamefont {He}, \citenamefont {Zhang}, \citenamefont {Tang},
  \citenamefont {Lin}, \citenamefont {Liu}, \citenamefont {Nie}, \citenamefont
  {Feng}, \citenamefont {Li}, \citenamefont {Xin}, \citenamefont {Ai},\ and\
  \citenamefont {Lu}}]{PhysRevLett.129.070502}%
  \BibitemOpen
  \bibfield  {author} {\bibinfo {author} {\bibfnamefont {X.}~\bibnamefont
  {Long}}, \bibinfo {author} {\bibfnamefont {W.-T.}\ \bibnamefont {He}},
  \bibinfo {author} {\bibfnamefont {N.-N.}\ \bibnamefont {Zhang}}, \bibinfo
  {author} {\bibfnamefont {K.}~\bibnamefont {Tang}}, \bibinfo {author}
  {\bibfnamefont {Z.}~\bibnamefont {Lin}}, \bibinfo {author} {\bibfnamefont
  {H.}~\bibnamefont {Liu}}, \bibinfo {author} {\bibfnamefont {X.}~\bibnamefont
  {Nie}}, \bibinfo {author} {\bibfnamefont {G.}~\bibnamefont {Feng}}, \bibinfo
  {author} {\bibfnamefont {J.}~\bibnamefont {Li}}, \bibinfo {author}
  {\bibfnamefont {T.}~\bibnamefont {Xin}}, \bibinfo {author} {\bibfnamefont
  {Q.}~\bibnamefont {Ai}},\ and\ \bibinfo {author} {\bibfnamefont
  {D.}~\bibnamefont {Lu}},\ }\href
  {https://doi.org/10.1103/PhysRevLett.129.070502} {\bibfield  {journal}
  {\bibinfo  {journal} {Phys. Rev. Lett.}\ }\textbf {\bibinfo {volume} {129}},\
  \bibinfo {pages} {070502} (\bibinfo {year} {2022})}\BibitemShut {NoStop}%
\bibitem [{\citenamefont {Xin}\ \emph {et~al.}(2021)\citenamefont {Xin},
  \citenamefont {Che}, \citenamefont {Xi}, \citenamefont {Singh}, \citenamefont
  {Nie}, \citenamefont {Li}, \citenamefont {Dong},\ and\ \citenamefont
  {Lu}}]{PhysRevLett.126.110502}%
  \BibitemOpen
  \bibfield  {author} {\bibinfo {author} {\bibfnamefont {T.}~\bibnamefont
  {Xin}}, \bibinfo {author} {\bibfnamefont {L.}~\bibnamefont {Che}}, \bibinfo
  {author} {\bibfnamefont {C.}~\bibnamefont {Xi}}, \bibinfo {author}
  {\bibfnamefont {A.}~\bibnamefont {Singh}}, \bibinfo {author} {\bibfnamefont
  {X.}~\bibnamefont {Nie}}, \bibinfo {author} {\bibfnamefont {J.}~\bibnamefont
  {Li}}, \bibinfo {author} {\bibfnamefont {Y.}~\bibnamefont {Dong}},\ and\
  \bibinfo {author} {\bibfnamefont {D.}~\bibnamefont {Lu}},\ }\href
  {https://doi.org/10.1103/PhysRevLett.126.110502} {\bibfield  {journal}
  {\bibinfo  {journal} {Phys. Rev. Lett.}\ }\textbf {\bibinfo {volume} {126}},\
  \bibinfo {pages} {110502} (\bibinfo {year} {2021})}\BibitemShut {NoStop}%
\bibitem [{\citenamefont {Cheng}\ \emph {et~al.}(2023)\citenamefont {Cheng},
  \citenamefont {Deng}, \citenamefont {Gu}, \citenamefont {He}, \citenamefont
  {Hu}, \citenamefont {Huang}, \citenamefont {Li}, \citenamefont {Lin},
  \citenamefont {Lu}, \citenamefont {Lu} \emph {et~al.}}]{cheng2023noisy}%
  \BibitemOpen
  \bibfield  {author} {\bibinfo {author} {\bibfnamefont {B.}~\bibnamefont
  {Cheng}}, \bibinfo {author} {\bibfnamefont {X.-H.}\ \bibnamefont {Deng}},
  \bibinfo {author} {\bibfnamefont {X.}~\bibnamefont {Gu}}, \bibinfo {author}
  {\bibfnamefont {Y.}~\bibnamefont {He}}, \bibinfo {author} {\bibfnamefont
  {G.}~\bibnamefont {Hu}}, \bibinfo {author} {\bibfnamefont {P.}~\bibnamefont
  {Huang}}, \bibinfo {author} {\bibfnamefont {J.}~\bibnamefont {Li}}, \bibinfo
  {author} {\bibfnamefont {B.-C.}\ \bibnamefont {Lin}}, \bibinfo {author}
  {\bibfnamefont {D.}~\bibnamefont {Lu}}, \bibinfo {author} {\bibfnamefont
  {Y.}~\bibnamefont {Lu}}, \emph {et~al.},\ }\href
  {https://doi.org/10.1007/s11467-022-1249-z} {\bibfield  {journal} {\bibinfo
  {journal} {Front. Phys.}\ }\textbf {\bibinfo {volume} {18}},\ \bibinfo
  {pages} {21308} (\bibinfo {year} {2023})}\BibitemShut {NoStop}%
\bibitem [{\citenamefont {Baumgratz}\ \emph {et~al.}(2014)\citenamefont
  {Baumgratz}, \citenamefont {Cramer},\ and\ \citenamefont
  {Plenio}}]{Baumgratz2014quantifying}%
  \BibitemOpen
  \bibfield  {author} {\bibinfo {author} {\bibfnamefont {T.}~\bibnamefont
  {Baumgratz}}, \bibinfo {author} {\bibfnamefont {M.}~\bibnamefont {Cramer}},\
  and\ \bibinfo {author} {\bibfnamefont {M.~B.}\ \bibnamefont {Plenio}},\
  }\href {https://doi.org/10.1103/PhysRevLett.113.140401} {\bibfield  {journal}
  {\bibinfo  {journal} {Phys. Rev. Lett.}\ }\textbf {\bibinfo {volume} {113}},\
  \bibinfo {pages} {140401} (\bibinfo {year} {2014})}\BibitemShut {NoStop}%
\bibitem [{\citenamefont {Winter}\ and\ \citenamefont
  {Yang}(2016)}]{Winter2016operational}%
  \BibitemOpen
  \bibfield  {author} {\bibinfo {author} {\bibfnamefont {A.}~\bibnamefont
  {Winter}}\ and\ \bibinfo {author} {\bibfnamefont {D.}~\bibnamefont {Yang}},\
  }\href {https://doi.org/10.1103/PhysRevLett.116.120404} {\bibfield  {journal}
  {\bibinfo  {journal} {Phys. Rev. Lett.}\ }\textbf {\bibinfo {volume} {116}},\
  \bibinfo {pages} {120404} (\bibinfo {year} {2016})}\BibitemShut {NoStop}%
\bibitem [{\citenamefont {Lostaglio}\ \emph
  {et~al.}(2015{\natexlab{b}})\citenamefont {Lostaglio}, \citenamefont
  {Korzekwa}, \citenamefont {Jennings},\ and\ \citenamefont
  {Rudolph}}]{Lostaglio2015quantum}%
  \BibitemOpen
  \bibfield  {author} {\bibinfo {author} {\bibfnamefont {M.}~\bibnamefont
  {Lostaglio}}, \bibinfo {author} {\bibfnamefont {K.}~\bibnamefont {Korzekwa}},
  \bibinfo {author} {\bibfnamefont {D.}~\bibnamefont {Jennings}},\ and\
  \bibinfo {author} {\bibfnamefont {T.}~\bibnamefont {Rudolph}},\ }\href
  {https://doi.org/10.1103/PhysRevX.5.021001} {\bibfield  {journal} {\bibinfo
  {journal} {Phys. Rev. X}\ }\textbf {\bibinfo {volume} {5}},\ \bibinfo {pages}
  {021001} (\bibinfo {year} {2015}{\natexlab{b}})}\BibitemShut {NoStop}%
\bibitem [{\citenamefont {\ifmmode \acute{C}\else
  \'{C}\fi{}wikli\ifmmode~\acute{n}\else \'{n}\fi{}ski}\ \emph
  {et~al.}(2015)\citenamefont {\ifmmode \acute{C}\else
  \'{C}\fi{}wikli\ifmmode~\acute{n}\else \'{n}\fi{}ski}, \citenamefont
  {Studzi\ifmmode~\acute{n}\else \'{n}\fi{}ski}, \citenamefont {Horodecki},\
  and\ \citenamefont {Oppenheim}}]{cwiklinski2015limitations}%
  \BibitemOpen
  \bibfield  {author} {\bibinfo {author} {\bibfnamefont {P.}~\bibnamefont
  {\ifmmode \acute{C}\else \'{C}\fi{}wikli\ifmmode~\acute{n}\else
  \'{n}\fi{}ski}}, \bibinfo {author} {\bibfnamefont {M.}~\bibnamefont
  {Studzi\ifmmode~\acute{n}\else \'{n}\fi{}ski}}, \bibinfo {author}
  {\bibfnamefont {M.}~\bibnamefont {Horodecki}},\ and\ \bibinfo {author}
  {\bibfnamefont {J.}~\bibnamefont {Oppenheim}},\ }\href
  {https://doi.org/10.1103/PhysRevLett.115.210403} {\bibfield  {journal}
  {\bibinfo  {journal} {Phys. Rev. Lett.}\ }\textbf {\bibinfo {volume} {115}},\
  \bibinfo {pages} {210403} (\bibinfo {year} {2015})}\BibitemShut {NoStop}%
\bibitem [{\citenamefont {Toyabe}\ \emph {et~al.}(2010)\citenamefont {Toyabe},
  \citenamefont {Sagawa}, \citenamefont {Ueda}, \citenamefont {Muneyuki},\ and\
  \citenamefont {Sano}}]{toyabe2010information}%
  \BibitemOpen
  \bibfield  {author} {\bibinfo {author} {\bibfnamefont {S.}~\bibnamefont
  {Toyabe}}, \bibinfo {author} {\bibfnamefont {T.}~\bibnamefont {Sagawa}},
  \bibinfo {author} {\bibfnamefont {M.}~\bibnamefont {Ueda}}, \bibinfo {author}
  {\bibfnamefont {E.}~\bibnamefont {Muneyuki}},\ and\ \bibinfo {author}
  {\bibfnamefont {M.}~\bibnamefont {Sano}},\ }\href
  {https://doi.org/10.1038/nphys1821} {\bibfield  {journal} {\bibinfo
  {journal} {Nat. Phys.}\ }\textbf {\bibinfo {volume} {6}},\ \bibinfo {pages}
  {988} (\bibinfo {year} {2010})}\BibitemShut {NoStop}%
\bibitem [{\citenamefont {Sagawa}\ and\ \citenamefont
  {Ueda}(2008)}]{sagawa2008second}%
  \BibitemOpen
  \bibfield  {author} {\bibinfo {author} {\bibfnamefont {T.}~\bibnamefont
  {Sagawa}}\ and\ \bibinfo {author} {\bibfnamefont {M.}~\bibnamefont {Ueda}},\
  }\href {https://doi.org/10.1103/PhysRevLett.100.080403} {\bibfield  {journal}
  {\bibinfo  {journal} {Phys. Rev. Lett.}\ }\textbf {\bibinfo {volume} {100}},\
  \bibinfo {pages} {080403} (\bibinfo {year} {2008})}\BibitemShut {NoStop}%
\bibitem [{\citenamefont {Perarnau-Llobet}\ \emph {et~al.}(2015)\citenamefont
  {Perarnau-Llobet}, \citenamefont {Hovhannisyan}, \citenamefont {Huber},
  \citenamefont {Skrzypczyk}, \citenamefont {Brunner},\ and\ \citenamefont
  {Ac\'{\i}n}}]{PhysRevX.5.041011}%
  \BibitemOpen
  \bibfield  {author} {\bibinfo {author} {\bibfnamefont {M.}~\bibnamefont
  {Perarnau-Llobet}}, \bibinfo {author} {\bibfnamefont {K.~V.}\ \bibnamefont
  {Hovhannisyan}}, \bibinfo {author} {\bibfnamefont {M.}~\bibnamefont {Huber}},
  \bibinfo {author} {\bibfnamefont {P.}~\bibnamefont {Skrzypczyk}}, \bibinfo
  {author} {\bibfnamefont {N.}~\bibnamefont {Brunner}},\ and\ \bibinfo {author}
  {\bibfnamefont {A.}~\bibnamefont {Ac\'{\i}n}},\ }\href
  {https://doi.org/10.1103/PhysRevX.5.041011} {\bibfield  {journal} {\bibinfo
  {journal} {Phys. Rev. X}\ }\textbf {\bibinfo {volume} {5}},\ \bibinfo {pages}
  {041011} (\bibinfo {year} {2015})}\BibitemShut {NoStop}%
\bibitem [{\citenamefont {Rio}\ \emph {et~al.}(2011)\citenamefont {Rio},
  \citenamefont {{\AA}berg}, \citenamefont {Renner}, \citenamefont {Dahlsten},\
  and\ \citenamefont {Vedral}}]{rio2011thermodynamic}%
  \BibitemOpen
  \bibfield  {author} {\bibinfo {author} {\bibfnamefont {L.~d.}\ \bibnamefont
  {Rio}}, \bibinfo {author} {\bibfnamefont {J.}~\bibnamefont {{\AA}berg}},
  \bibinfo {author} {\bibfnamefont {R.}~\bibnamefont {Renner}}, \bibinfo
  {author} {\bibfnamefont {O.}~\bibnamefont {Dahlsten}},\ and\ \bibinfo
  {author} {\bibfnamefont {V.}~\bibnamefont {Vedral}},\ }\href
  {https://doi.org/10.1038/nature10123} {\bibfield  {journal} {\bibinfo
  {journal} {Nature}\ }\textbf {\bibinfo {volume} {474}},\ \bibinfo {pages}
  {61} (\bibinfo {year} {2011})}\BibitemShut {NoStop}%
\bibitem [{\citenamefont {Tang}\ \emph {et~al.}(2024)\citenamefont {Tang},
  \citenamefont {Guo}, \citenamefont {Hu}, \citenamefont {Huang}, \citenamefont
  {Liu}, \citenamefont {Li},\ and\ \citenamefont
  {Guo}}]{Tang2024demonstration}%
  \BibitemOpen
  \bibfield  {author} {\bibinfo {author} {\bibfnamefont {H.}~\bibnamefont
  {Tang}}, \bibinfo {author} {\bibfnamefont {Y.}~\bibnamefont {Guo}}, \bibinfo
  {author} {\bibfnamefont {X.-M.}\ \bibnamefont {Hu}}, \bibinfo {author}
  {\bibfnamefont {Y.-F.}\ \bibnamefont {Huang}}, \bibinfo {author}
  {\bibfnamefont {B.-H.}\ \bibnamefont {Liu}}, \bibinfo {author} {\bibfnamefont
  {C.-F.}\ \bibnamefont {Li}},\ and\ \bibinfo {author} {\bibfnamefont {G.-C.}\
  \bibnamefont {Guo}},\ }\href@noop {} {\bibfield  {journal} {\bibinfo
  {journal} {arXiv}\ } (\bibinfo {year} {2024})}\BibitemShut {NoStop}%
\end{thebibliography}%

\end{document}